\documentclass[journal]{IEEEtran}%

\IEEEoverridecommandlockouts                              %

\usepackage{xr}

\title{%
Self-Configurable Mesh-Networks for Scalable Distributed Submodular Bandit Optimization 
}
\author{Zirui Xu, Vasileios Tzoumas$^\dagger$
	\thanks{%
 $^\dagger$Department of Aerospace Engineering, University of Michigan, Ann Arbor, MI 48109 USA;  {\tt\footnotesize \{ziruixu,vtzoumas\}@umich.edu}} 
    \thanks{This work was supported by NSF CAREER Award No. 2337412 and ARO Early Career Program Award W911NF-25-1-0280.}
}

\usepackage{cite}

\usepackage{comment}
\usepackage{siunitx}
\usepackage{relsize}
\usepackage{ifthen}
\usepackage[colorinlistoftodos]{todonotes}

\usepackage[vlined,ruled,linesnumbered]{algorithm2e}
\usepackage{graphics} %
\usepackage{rotating}
\usepackage{color}
\usepackage{enumerate}
\usepackage[T1]{fontenc}
\usepackage{psfrag}
\usepackage{epsfig} %
\usepackage{booktabs}
\usepackage{graphicx,url}
\usepackage{multirow}
\usepackage{array}
\usepackage{latexsym}
\usepackage{amsfonts}
\usepackage{amsmath}
\usepackage{amssymb}
\usepackage{xstring}
\usepackage{algorithmic}
\usepackage{multirow}
\usepackage{xcolor}
\usepackage{prettyref}
\usepackage{flexisym}
\usepackage{bigdelim}
\usepackage{breqn} %
\usepackage{listings}
\let\labelindent\relax
\usepackage{xspace}
\xspaceaddexceptions{.,;:!?)'}

\usepackage{bm}
\graphicspath{{./figures/}}
\usepackage{tikz}
\usetikzlibrary{matrix,calc}
\usepackage{lipsum}
\usepackage{mdwlist}

\let\itemize\undefined
\makecompactlist{itemize}{stditemize}
\usepackage{enumitem}
\usepackage{caption}

\usepackage{amsthm}
\usepackage{mathtools}

\makeatletter
\let\NAT@parse\undefined
\makeatother
\usepackage{hyperref}
\usepackage{cleveref}
\usepackage{mleftright}

\hypersetup{%
  colorlinks=true,
  linkcolor=blue,
  filecolor=magenta,      
  urlcolor=black,
  citecolor=red,
  linkbordercolor={0 0 1}
}

\newtheorem{theorem}{Theorem}
\newtheorem{problem}{Problem}

\newtheorem{corollary}{Corollary}

\newtheorem{lemma}{Lemma}

\newtheorem{definition}{Definition}
\newtheorem{proposition}{Proposition}

\newcommand{\bdmath}{\begin{dmath}}
\newcommand{\edmath}{\end{dmath}}
\newcommand{\beq}{\begin{equation}}
\newcommand{\eeq}{\end{equation}}
\newcommand{\bdm}{\begin{displaymath}}
\newcommand{\edm}{\end{displaymath}}
\newcommand{\bea}{\begin{eqnarray}}
\newcommand{\eea}{\end{eqnarray}}
\newcommand{\beal}{\beq \begin{array}{lll}}
\newcommand{\eeal}{\end{array} \eeq}
\newcommand{\beas}{\begin{eqnarray*}}
\newcommand{\eeas}{\end{eqnarray*}}
\newcommand{\ba}{\begin{array}}
\newcommand{\ea}{\end{array}}
\newcommand{\bit}{\begin{itemize}}
\newcommand{\eit}{\end{itemize}}
\newcommand{\ben}{\begin{enumerate}}
\newcommand{\een}{\end{enumerate}}

\newcommand{\calA}{{\cal A}}
\newcommand{\calB}{{\cal B}}
\newcommand{\calC}{{\cal C}}

\newcommand{\calE}{{\cal E}}
\newcommand{\calF}{{\cal F}}
\newcommand{\calG}{{\cal G}}
\newcommand{\calH}{{\cal H}}

\newcommand{\calJ}{{\cal J}}

\newcommand{\calM}{{\cal M}}
\newcommand{\calN}{{\cal N}}

\newcommand{\calV}{{\cal V}}

\definecolor{myblue}{RGB}{65 105 225}

\newcommand{\hide}[1]{}

\newcommand{\hiddenText}{{\color{gray} hidden text.}}
\newcommand{\hideWithText}[1]{\hiddenText}

\newcommand{\opt}{^{\star}}

\DeclareRobustCommand{\scenario}[1]{%
  \ifmmode
    \mathsf{#1}%
  \else
    \texorpdfstring{{\fontsize{8.9}{9}\selectfont\sc #1}\xspace}{#1}%
  \fi
}

\newcommand{\ie}{\emph{i.e.},\xspace}
\newcommand{\eg}{\emph{e.g.},\xspace}
\newcommand{\myin}{\, \in \,}

\newcommand{\blue}[1]{{\color{blue}#1}}

\newcommand{\sg}{\scenario{SG}}
\newcommand{\banalg}{\scenario{BSG}}

\newcommand{\dfs}{\scenario{DFS-SG}}

\newcommand{\myParagraph}[1]{{\bf #1.}\xspace}

\renewcommand{\opt}{\fontsize{9}{9}\selectfont \sf OPT}

\newcommand{\curv}{\kappa}

\newcommand{\alg}{\scenario{Anaconda}}
\newcommand{\actionsel}{\scenario{ActSel}}
\newcommand{\neighborsel}{\scenario{NeiSel}}

\newcommand{\elem}{v}

\newcommand{\distfsf}{p}

\newcommand{\solopt}{\calA^{\opt}}

\newcommand{\expthree}{\scenario{Exp3}}

\newcommand{\AReg}{\operatorname{A-Reg}_{T}}

\newcommand{\expthreeix}{\scenario{Exp3-IX}}
\newcommand{\smi}[2]{\text{{\fontsize{9}{9}\selectfont\sf VoC}}_{f,t}\!\left({#1};\,{#2}\right)}
\newcommand{\voc}{{\fontsize{9}{9}\selectfont\sf VoC}}

\PassOptionsToPackage{end}{algorithmic}

\begin{document}

\maketitle

\thispagestyle{empty}
\pagestyle{empty}

\begin{abstract}
We study how to scale distributed bandit submodular coordination under realistic communication constraints in bandwidth, data rate, and connectivity. We are motivated by multi-agent tasks of active situational awareness in unknown, partially-observable, and resource-limited environments, where the agents must coordinate through agent-to-agent communication.
Our approach enables scalability by (i) limiting information relays to only one-hop communication and (ii) keeping inter-agent messages small, having each agent transmit only its own action information. Despite these information-access restrictions, our approach enables near-optimal action coordination by optimizing the agents' communication neighborhoods over time, through distributed online bandit optimization, subject to the agents' bandwidth constraints.
Particularly, our approach
enjoys an anytime suboptimality bound that is also strictly positive for arbitrary network topologies, even disconnected.
To prove the bound, we define the \emph{Value of Coordination} ({\fontsize{8}{8}\selectfont\sf VoC}), an information-theoretic
metric that quantifies for each agent the benefit of information access to its neighbors. 
We validate in simulations the scalability and near-optimality of our approach:
it is observed to converge faster, outperform benchmarks for bandit submodular coordination, and can even outperform benchmarks that are privileged with a priori knowledge of the environment.
\end{abstract}

\vspace{-3.3mm}
\section{Introduction}\label{sec:Intro}

In the future, large-scale teams of distributed agents will be executing sensing-driven tasks such as {target tracking}~\cite{xu2023bandit}, {environmental mapping}~\cite{atanasov2015decentralized}, and {area monitoring}~\cite{corah2018distributed}. 
These collaborative tasks require the distributed agents to share their local observations---expand information access---to improve coordination performance.
However, expanding information access is challenging due to the limited  communication, computing, and sensing capabilities onboard each agent, 
and the above tasks often operate in 
unknown, unstructured, and dynamic environments.
In more detail, several challenges to scalability and optimality appear: 

    \textit{(a) Limited communication bandwidth:} Each agent can communicate with only a limited number of peers at a time, rather than with all physically reachable ones~\cite{jadbabaie2003coordination,nedic2009distributed}.
    
    \textit{(b) Finite data rate:} Communication occurs via onboard radio modules with restricted data rates, ranging from as low as 0.25 Mbps (\eg Digi XBee 3 Zigbee 3~\cite{digi_xbee3_zigbee3}) to around 100 Mbps (\eg Silvus Tech SL5200~\cite{silvus_streamcaster_5200}), far below the 0.8--1.5 Gbps commonly available in everyday Wi-Fi 6 systems~\cite{Tachus_WiFi6_vs_WiFi6E_2025}. %
    
    \textit{(c) Unguaranteed network connectivity:} Agents may receive information only from those within a fixed communication range or line of sight, meaning the communication graph is not guaranteed to remain connected~\cite{kantaros2019temporal}.

    \textit{(d) Unstructured environment:} The environment may be unknown a priori 
and partially observable.  Thus, the informational value of each observation becomes known only once the observation is made. 
This optimization setting corresponds to the \textit{bandit} feedback model~\cite{lattimore2020bandit}.

\begin{figure*}[t]
\captionsetup{font=footnotesize}
    \centering
\hspace{0.4mm}\includegraphics[width=.98\textwidth]{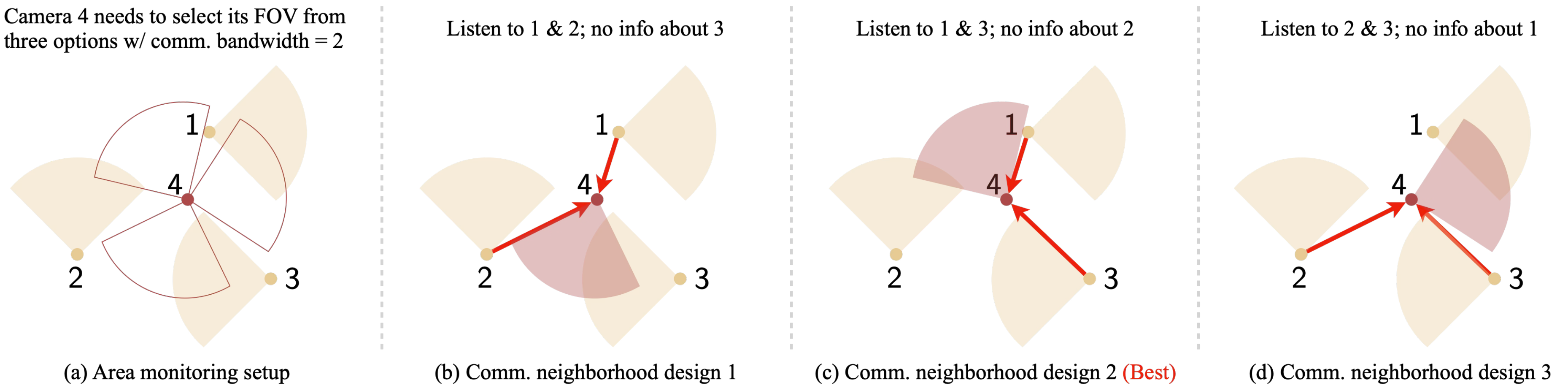}
    \vspace{-1mm}
    \caption{\textbf{Information Access Matters: A Multi-Camera Area Monitoring Example.} 
    Consider a multi-camera area monitoring task where four cameras must coordinate their fields of view (FOVs) via distributed communication to maximize total coverage. As shown in (a), suppose that cameras 1--3 have already fixed their FOVs (soft orange), and camera 4 must select its FOV from three predefined options (dark red). While the optimal choice for camera~4 depends on the FOVs of all other three, its communication bandwidth allows it to receive information from at most two of them at any current time. The three possible communication neighborhood configurations and the corresponding FOV selections are demonstrated in (b)--(d), among which the design in (c) yields the highest coverage and therefore the optimal FOV decision. This example illustrates that intelligent information access—enabled by active neighborhood design (possibly over multiple time steps)—can optimize action coordination performance in distributed settings with limited communication resources.
    }\label{fig:intro}
   \vspace{-5mm}
\end{figure*}

These challenges motivate distributed optimization problems in bandit settings~\cite{lattimore2020bandit} for \textit{large-scale} tasks in robotics, control, and machine learning. Such joint optimization problems often take the form of
\begin{equation}\label{eq:intro-online}
\max_{a_{i,t}\myin\mathcal{V}_i, \forall\, i\myin \calN,\, \forall\, t\myin [T]}\;
\sum_{t=1}^T\,
f\bigl(\,\{a_{i,t}\}_{i\myin \calN}\,\bigr),
\end{equation}
where $T$ is the operation time-horizon, and at each time step $t\in[T]$, the agents $\calN$ need to each select an action $a_{i,t}$ from its available action set $\calV_i$ to collaboratively maximize the unknown a priori objective function $f\colon \prod_{i \in \calN}\calV_i \mapsto\mathbb{R}$ that captures the task utility~\cite{krause2008near,singh2009efficient,tokekar2014multi,atanasov2015decentralized,gharesifard2017distributed,marden2017role,grimsman2019impact,corah2018distributed,schlotfeldt2021resilient,du2022jacobi,rezazadeh2023distributed,robey2021optimal,xu2025communication}. In information-gathering tasks, $f$ is often \textit{submodular}~\cite{fisher1978analysis}: submodularity is a diminishing returns property, and it emanates due to the possible information overlap among the information gathered by the agents~\cite{krause2008near}. For example, in target monitoring with multiple cameras at fixed locations, $\calN$ is the set of cameras,  $\calV_i$ is the available directions the camera can choose, and $f$ is the number of targets covered by the cameras' collective field of view (FOV). The cameras have no prior knowledge of target locations and can observe them only when they fall within the cameras’ collective FOV; targets that remain uncovered are therefore unknown to the cameras. As a result, the cameras can accurately evaluate only the utility of the FOVs they actually choose, which corresponds to the bandit feedback setting~\cite{lattimore2020bandit}.

The optimization problem in~\cref{eq:intro-online} is NP-hard even in known environments~\cite{Feige:1998:TLN:285055.285059}.  Polynomial-time algorithms with provable approximation guarantees exist. A classical example is the Sequential Greedy (\sg) algorithm~\cite{fisher1978analysis}, which achieves a $1/2$-approximation ratio. Since many multi-agent tasks, such as target tracking, collaborative mapping, and area monitoring, can be formulated as submodular coordination problems, \sg and its variants have been widely adopted across the controls, machine learning, and robotics communities~\cite{krause2008near,singh2009efficient,tokekar2014multi,atanasov2015decentralized,gharesifard2017distributed,grimsman2019impact,corah2018distributed,schlotfeldt2021resilient,liu2021distributed,robey2021optimal,rezazadeh2023distributed,konda2022execution,krause2012submodular,xu2025communication}. In unknown environments instead, online variants of \sg are proposed~\cite{streeter2008online,streeter2009online,suehiro2012online,golovin2014online,chen2018online,zhang2019online,xu2023online,xu2023bandit}, providing guaranteed suboptimality over the horizon $T$. 
Continuous-domain optimization techniques have also been leveraged in the bandit setting by first using the multilinear extension~\cite{calinescu2011maximizing}, to lift the discrete submodular function $f$ to the continuous domain, then applying gradient/consensus-based techniques, and finally rounding back to the discrete domain to obtain a near-optimal solution~\cite{mokhtari2018decentralized,zhang2019online,chen2020black,zhang2025near,robey2021optimal}.

However, the algorithms above cannot scale as the network grows when challenges (a)--(d) exist simultaneously. %
In particular, current distributed approaches, based on either sequential communication in the discrete domain~\cite{atanasov2015decentralized,xu2023online,xu2023bandit} or repeated consensus iterations in the continuous domain~\cite{streeter2008online,streeter2009online,suehiro2012online,golovin2014online,chen2018online,mokhtari2018decentralized,zhang2019online,chen2020black,robey2021optimal,liu2021distributed,zhang2025near}, assume instantaneous agent-to-agent communication, \ie infinite data rates. But if accounting for the limited data rates of onboard radio modules, per challenge (b), decision times of these methods can scale cubically in network size or even more~\cite[Table 1]{xu2025communication}. This is due to: 
\begin{itemize}[leftmargin=*]
    \item \textit{Message size:} The size of communication messages can scale proportionally with the number of agents. %
    \item \textit{Information relay:} Distributed agents use multi-hop relays to expand information access. %
\end{itemize}

In sum, under challenges (a)--(d), where instantaneous communication is infeasible and the environment is unknown, achieving both scalability and optimality requires \textit{restricting information access intelligently}, via limiting message size and/or information relay, to balance the trade-off between decision time and decision optimality. Methods in~\cite{rikos2023asynchronous,hu2025distributed} leverage communication quantization to shorten messages and evaluate its effect on optimality. 
The framework in~\cite{bianchi2024end} reduces communication overhead by exploiting the sparsity of the underlying information dependence among agents and omitting information sharing between agents with low dependence accordingly. It also establishes the necessary conditions for convergence. 
However, the approaches in~\cite{rikos2023asynchronous,hu2025distributed,bianchi2024end}
focus on convex optimization. %
In discrete submodular optimization, the focus of this paper, current works~\cite{gharesifard2017distributed,marden2017role,grimsman2019impact} only perform $f$-agnostic communication restriction, \ie the topology designed will remain the same, independently of the $f$ specified by the application.
In more detail,~\cite{gharesifard2017distributed,marden2017role,grimsman2019impact} study a variant of \sg with restricted information relay where each agent $i$ receives information from only a subset rather than all of $\{1,\dots,i-1\}$ as in the classical \sg.\footnote{In discrete submodular optimization, agents perform sequential communication, where message complexity depends on the network topology. For example, in a line network, a message from agent $i-1$ to $i$ may aggregate information from all preceding agents $\{1,\dots,i-1\}$, whereas in a complete network it contains only agent $i-1$’s data, with other information transmitted through separate links.} These works derive suboptimality bounds that scale inversely with the independence number of the resulting information-sharing topology,\footnote{This information-sharing topology is the Directed Acyclic Graph (DAG) derived from the communication network.} and further propose a centralized information-sharing topology design method based on these bounds~\cite{grimsman2019impact}.
The recent algorithm in~\cite{xu2025communication} restricts information access within one-hop neighbors, thus prohibiting relays and having each agent-to-agent message to contain information only about the agent that transmits it. It also enables partially parallelized action selection to reduce communication latency. Although~\cite{xu2025communication}  characterizes $f$-specific coordination performance, the bound is intractable to optimize even in hindsight, and therefore cannot support intelligent information access restriction.

\myParagraph{Contributions}
We provide a distributed multi-agent decision-making framework that enables both scalable and near-optimal action coordination in unknown environments under realistic communication constraints in bandwidth, data rate, and connectivity. 
Our approach enables scalability by (i) limiting information relays to only one-hop communication and (ii) keeping inter-agent messages small, having each agent transmit only its own action information. Despite these information-access restrictions, our approach improves the near-optimality bound of action coordination by optimizing the agents' communication neighborhoods over time, through distributed online bandit optimization (subject to the agents' bandwidth constraints)---the necessity for such information access optimization is demonstrated in~Fig.~\ref{fig:intro}.

To our knowledge, this is the first rigorous approach that enables multi-agent networks to scale near-optimal coordination by actively optimizing the restricted information access through active communication topology self-configuration. %
The approach is fully distributed: each agent jointly selects its action and designs its local communication neighborhood subject to its coordination neighborhood and bandwidth constraints. %
The algorithm has the following properties:
\paragraph{Scalability} 
\alg enjoys improved scalability under communication constraints by having a convergence time of 
$O(|\calN|^2)$ in sparse networks, accounting for information relay in multi-hop communication (\Cref{sec:resource-guarantees}). This is faster than existing methods even though they only consider known environments~\cite[Table I]{xu2025communication}. In practice, the algorithm can convergence even faster, \eg only sublinearly in $|\calN|$ in the area monitoring simulations (\Cref{subsec:scalability}). 
\paragraph{Anytime Self-Configuration with Arbitrary Topologies}
\alg enables each agent to adapt its communication neighborhood online according to the given $f$, resulting in coordination over a directed, time-varying, and potentially disconnected network. The fully distributed self-configuration mechanism further allows agents to join or leave the system without disrupting the co-optimization process. Our prior work \cite{xu2025communication}, instead, does not optimize the network; \cite{grimsman2019impact} performs network design centrally without leveraging $f$; and \cite{atanasov2015decentralized,corah2018distributed,schlotfeldt2021resilient,liu2021distributed,robey2021optimal,rezazadeh2023distributed,konda2022execution,xu2023bandit,xu2023online} require connected networks.
\paragraph{Approximation Performance} 
\alg enjoys anytime, strictly positive, $f$-specific suboptimality bounds against an optimal solution of \cref{eq:intro-online} (\Cref{sec:suboptimality-guarantees}). 
The bounds capture the benefit of information access for each agent and thus remain valid under arbitrarily optimized communication topologies, regardless of global connectivity. 
In particular, \Cref{th:main}, along with our numerical evaluations, validates that coordination performance improves through network optimization, %
and \Cref{th:posteriori,th:asymptotic} ensure that \alg achieves strictly positive performance guarantees at all times. The bounds in~\cite{grimsman2019impact,gharesifard2017distributed,marden2017role} are instead $f$-agnostic; %
the bound in our prior work~\cite{xu2025communication} does not support network optimization; and \cite{atanasov2015decentralized,corah2018distributed,schlotfeldt2021resilient,liu2021distributed,robey2021optimal,rezazadeh2023distributed,konda2022execution,xu2023bandit,xu2023online} provide suboptimality guarantees only under connected network assumptions.

\myParagraph{Numerical evaluations}
We validate \alg through simulations of multi-camera area monitoring. The results first show that the proposed information-driven neighbor selection strategy  outperforms typical baselines (nearest and random neighbors) in robotics and controls~\cite{zhou2023racer,liu2025slideslam,xu2025communication}, with particularly large gaps in certain structured environments (\Cref{sec:experiments-1}). Then, comparisons with state-of-the-art centralized and sequential methods (\scenario{DFS-SG} and \scenario{DFS-BSG}) under both idealized and realistic delay settings reveal a fundamental trade-off between coordination optimality and convergence speed, and highlight the importance of delay-aware evaluation for real-time performance. Although the benchmarks require relaxed versions of the problem in~\cref{eq:intro-online} to be applicable, \alg still achieves competitive or better coverage (\Cref{subsec:compared-algorithms,subsec:comparison-no-delay,subsec:comparison-with-delays}). Finally, large-scale simulations confirm the scalability of \alg: while benchmarks such as \scenario{DFS-BSG} incur prohibitive communication delays, \alg maintains a constant time per decision round and can scale sublinearly in network size in practice (\Cref{subsec:scalability}). %
{We ran all simulations using Python 3.11.7 on a Windows PC with an Intel Core i9-14900KF CPU @ 3.20 GHz and 64 GB RAM. 
The code is available at \href{https://github.com/UM-iRaL/Self-configurable-network}{\blue{https://github.com/UM-iRaL/Self-configurable-network}}.}

\myParagraph{Comparison with preliminary work~\cite{xu2024performance,xu2025communication}} This paper extends our preliminary work~\cite{xu2024performance} to include new theoretical results, extensive evaluations, and proofs of all claims. We introduce a tighter a priori bound (\Cref{th:main}), a strictly positive a posteriori bound (\Cref{th:posteriori}), and a strictly positive asymptotic bound (\Cref{th:asymptotic}), and we comprehensively evaluate the proposed algorithm across coordination performance, neighborhood design, and scalability metrics in area monitoring scenarios (\Cref{sec:experiments-1}).
This paper also extends the formulation in prior work~\cite{xu2025communication} where communication neighborhoods are fixed and constructed heuristically (\eg via nearest neighbors). Here, we perform network optimization and demonstrate that the resulting configurations outperform nearest neighbors in area monitoring scenarios (\Cref{sec:experiments-1}).

\section{Distributed Coordination and Network Co-Optimization}\label{sec:problem}

We present the problem formulation of \textit{Distributed Coordination and Network Co-Optimization}. We use the notation:
\begin{itemize}[leftmargin=*]
    \item $\calV_\calN \triangleq \prod_{i\myin \calN} \,\calV_i$ 
    is the cross product of sets $\{\calV_i\}_{i\myin \calN}$.
    \item $[T]\triangleq\{1,\dots,T\}$ for any positive integer $T$.
    \item $f(\,a\,|\,\calA\,)\triangleq f(\,\calA \cup \{a\}\,)-f(\,\calA\,)$ is the marginal gain of set function $f:2^\calV\mapsto \mathbb{R}$ for adding $a \in \calV$ to $\calA \subseteq\calV$.
    \item $|\calA|$ is the cardinality of a discrete set $\calA$. 
\end{itemize}
We also lay down the following framework about the agents' communication network and the objective function $f$.

\myParagraph{Communication network} The agents' communication network $\calG_t=(\calN, \calE_t)$, $\forall t$ is \textit{undetermined} a priori, where $\calE_t$ is the set of (directed) communication edges among agents $\calN$ at time $t$.  The goal of this work is for the agents to jointly optimize $\calE_t$ to enable convergence to a near-optimal solution to \eqref{eq:intro-online} despite distributed coordination. The network optimized by $\calN$ \textit{can be directed and even disconnected}. We refer to the case where every agent receives information from all others as \textit{fully centralized}, and the case where no agent receives information from any other as \textit{fully decentralized}.

\myParagraph{Communication neighborhood}  
When a communication channel exists from agent $j$ to agent $i$ at time $t$, \ie $(j\rightarrow i) \in \calE_t$, then $i$ can receive, store, and process information sent by $j$, and the set of all such $j$ is $i$'s \textit{communication neighborhood}, denoted as $\calN_{i,t}$.   

\myParagraph{Communication constraints} Each agent $i$ can receive information from up to $\alpha_i$ other agents at the same time due to onboard bandwidth constraints, \ie $|\calN_{i,t}|\;\leq \alpha_i$, $\forall t$. 
Also, we denote as $\calM_i\subseteq \calN\setminus\{i\}$ the set of agents that can potentially send information to agent $i$ ---not all $\calN\setminus\{i\}$ can reach agent $i$ due to distance or obstacles: agent $i$ can pick its neighbors by choosing at most $\alpha_i$ agents from $\calM_i$. We refer to $\calM_i$ as agent $i$'s \textit{coordination neighborhood} ($\calN_{i,t}\subseteq\calM_i$, $\forall t$).

\begin{definition}[Normalized and Non-Decreasing Submodular Set Function{~\cite{fisher1978analysis}}]\label{def:submodular}
A set function $f:2^\calV\mapsto \mathbb{R}$ is \emph{normalized and non-decreasing submodular} if and only if 
\begin{itemize}[leftmargin=*]
\item (Normalization) $f(\emptyset)=0$;
\item (Monotonicity) $f(\calA)\leq f(\calB)$, $\forall\,\calA\subseteq \calB\subseteq \calV$;
\item (Submodularity) $f(s\,|\,\calA)\geq f(s\,|\,\calB)$, $\forall\,\calA\subseteq {\mathcal{B}}\subseteq\calV$ and $s\in \calV$.
\end{itemize}
\end{definition}

Intuitively, if $f(\calA)$ captures the number of targets tracked by a set $\calA$ of sensors, then the more sensors are deployed, more or the same targets are covered; this is the non-decreasing property.  Also, the marginal gain of tracked targets caused by deploying a sensor $s$ \emph{drops} when \emph{more} sensors are already deployed; this is the submodularity~property.

\begin{definition}[2nd-order Submodular Set Function{~\cite{crama1989characterization,foldes2005submodularity}}]\label{def:conditioning}
$f:2^\calV\mapsto \mathbb{R}$ is \emph{2nd-order submodular} if and only if 
\begin{equation}\label{eq:conditioning}
    f(s\,|\,\calC) - f(s\,|\,\calA\cup\calC) \geq f(s\,|\,\calB\cup\calC) - f(s\,|\,\calA\cup\calB\cup\calC),
\end{equation}
for any \emph{disjoint} $\calA, \calB, \calC\subseteq \calV$ ($\calA \cap \calB \cap \calC =\emptyset$) and  $s\in\calV$.
\end{definition}

Intuitively, if $f(\calA)$ captures the number of targets tracked by a set $\calA$ of sensors, then \emph{marginal gain of the marginal gains} drops when more sensors are already deployed.

\begin{problem}[Distributed  Coordination and Network Co-Optimization]
\label{pr:online}
At each $t\in[T]$, each agent $i\in\calN$ selects a communication neighborhood $\calN_{i,t}$ of size up to $\alpha_i$ and an action $a_{i,t}$ to solve the optimization problem
\begin{equation}\label{eq:problem}
\max_{\substack{\calN_{i,t}\,\subseteq\,\mathcal{M}_i,\, |\calN_{i,t}| \,\le\, \alpha_i \\ \forall\, i\myin \calN,\, \forall\, t\myin [T]}}\,
\max_{\substack{a_{i,t}\myin\mathcal{V}_i \\ \forall\, i\myin \calN,\, \forall\, t\myin [T]}}\;
\sum_{t=1}^T\,
f\bigl(\,\{a_{i,t}\}_{i\myin \calN}\,\bigr),
\end{equation}where (i) each agent $i$ can coordinate with its neighbors only, without any information about non-neighbors, (ii) $f\colon \calV_\calN \mapsto\mathbb{R}$ is a normalized, non-decreasing submodular, and 2nd-order submodular, and (iii) $f$ is known via bandit feedback, in particular, each agent $i$ can access the value of $f(\calA_t)$ only, $\forall\,\calA_t\subseteq \{a_{i,t}\}\cup\{a_{j,t}\}_{j\in \calN_{i,t}}$, once the agent has selected $a_{i,t}$ and received $\{a_{j,t}\}_{j\in \calN_{i,t}}$ from neighbors.
\end{problem}

\Cref{pr:online} captures the intrinsic coupling between coordination performance and information access, as illustrated by the example in \Cref{fig:intro}. The objective of this paper is to design the communication network topology that maximizes action coordination performance by solving \Cref{pr:online}.

\vspace{-1mm}\section{Alternating Coordination and Network-Design Algorithm (\alg)} \label{sec:algorithm}

\setlength{\textfloatsep}{3mm}
\begin{algorithm}[t]
	\caption{AlterNAting COordination and Network Design Algorithm (\alg) for Agent $i$
	}
	\begin{algorithmic}[1]
		\REQUIRE \!Time horizon $T$; agent $i$'s coordination neighborhood $\calM_i$; agent $i$'s communication bandwidth $\alpha_i$. %
		\ENSURE \!Action $a_{i, t}$ and communication neighborhood $\calN_{i, t}$, $\forall t\in[T]$.
		\medskip
        \STATE $\calN_{i,0}\gets\emptyset, \forall i \in \calN$; 
		\FOR{each time step $t\in [T]$}
        \STATE $a_{i,t}\gets\text{\actionsel}([T], \calV_i)$;
        \STATE $\calN_{i,t}\gets\text{\neighborsel}(a_{i,t}, [T], \calM_i, \alpha_i)$;
        \STATE \textbf{receive} neighbors' actions $\{a_{j, t}\}_{j\myin\calN_{i,t}}$ and \\\textbf{update} \actionsel (per lines 6-8) and \neighborsel (per lines 6-11);
		\ENDFOR
	\end{algorithmic}\label{alg:main}
\end{algorithm}

We present \alg. 
\alg approximates a solution to \Cref{pr:online} by alternating the optimization for action coordination and communication neighborhood design. %
Since both action coordination and communication neighborhood design take the form of adversarial bandit problems, we present first the adversarial bandit (\Cref{subsec:prelim}), then the algorithms \actionsel (\Cref{subsec:action}) and \neighborsel (\Cref{subsec:neighbor}).

\subsection{Adversarial Bandit Problem}\label{subsec:prelim}

The adversarial bandit problem involves an agent selecting a sequence of actions to maximize the total reward over a given number of time steps~\cite{lattimore2020bandit}.  The challenge is that, at each time step, no action's reward is known to the agent a priori, and after an action is selected, only the selected action's reward will become known. To present the problem, we have:
\begin{itemize}[leftmargin=*]
    \item $\calV$ denotes the available action set;
    \item $v_{t}\in\calV$ denotes the agent's selected action at time $t$;
    \item $r_{v_t,\,t}\in[0,1]$ denotes the reward of selecting $v_{t}$ at time $t$.
\end{itemize}

\begin{problem}[Adversarial Bandit~\cite{lattimore2020bandit}]\label{pr:bandit}
Assume a horizon of $T$ time steps. At each time step $t\in[T]$, the agent needs to select an action $v_t\in\calV$ such that the regret
\begin{equation}\label{eq:bandit}
    \operatorname{Regret}_T \triangleq\max_{v\myin\mathcal{V}} \, \sum_{t=1}^T\,r_{v,\,t} \,- \,\sum_{t=1}^T\,r_{v_t,\,t},
\end{equation}
is minimized, where no actions' rewards are known a priori, and only the reward $r_{v_t,\,t}\in[0,1]$ becomes known to the agent \emph{after} $v_t$ is selected. 

\end{problem}

The goal of solving \Cref{pr:bandit} is to achieve a sublinear $\operatorname{Regret}_T$, \ie $\operatorname{Regret}_T/T\rightarrow 0$ for $T\rightarrow \infty$, since this implies that the agent asymptotically chooses optimal actions even though the rewards are unknown a priori~\cite{lattimore2020bandit}. 
The most classical adversarial bandit algorithm \expthree~\cite{auer2002nonstochastic} %
achieves the goal above by obtaining a regret bound of $\Tilde{O}(\sqrt{|\calV|T})$, where $\Tilde{O}(\cdot)$ hides $\log$ terms. During the past two decades, several refined adversarial bandit algorithms have emerged, including \scenario{Exp3-IX} for high-probability regret bounds via implicit exploration~\cite{neu2015explore}, \scenario{Exp3++} for adaptation between stochastic and adversarial regimes~\cite{seldin2014one}, and \scenario{Tsallis-INF} for minimax-optimal regret with improved adaptivity through Tsallis entropy regularization~\cite{zimmert2021tsallis}.

In this paper, we will formulate both action coordination and neighbor selection problems in the form of \Cref{pr:bandit}, and employ \expthree as the bandit solver without loss of generality. Alternative methods such as those mentioned above are also applicable (\eg~\expthreeix was used in~\cite{xu2024performance}). %

\subsection{Action Coordination}\label{subsec:action}

\setlength{\textfloatsep}{3mm}
\begin{algorithm}[t]
	\caption{\actionsel for Agent $i$
	}
	\begin{algorithmic}[1]
		\REQUIRE \!Time horizon $T$ and agent $i$'s action set $\mathcal{V}_i$. %
		\ENSURE \!Agent $i$'s action $a_{i, t}$, $\forall t\in[T]$.
		\medskip
            \STATE $\eta_i^a\gets\sqrt{2\log{|\calV_i|}\,/\,(|\calV_i|T)}$;
            \STATE $w_{1}\gets\mleft[w_{1,1}, \dots, w_{|\calV_i|,1}\mright]^\top$ with $w_{v,1}=1, \forall a\in \calV_i$;
            \FOR {\text{each time step} $t\in [T]$}
		\STATE \textbf{get} distribution $\distfsf_t\gets{w_t}\,/\,{\|w_t\|_1}$; 
		\STATE \textbf{draw} action $a_{i,t}\in\calV_i$ \textbf{from} $\distfsf_t$;
		\STATE \textbf{input} $a_{i,t}$ \textbf{to} \neighborsel and \\
        \textbf{receive} neighbors' actions $\{a_{j, t}\}_{j\in\calN_{i,t}}$;
		\STATE $r_{a_{i,t}, t}\gets f(\,a_{i,t}\,|\, \{a_{j, t}\}_{j\in\calN_{i,t}}\,)$ and \\
  \textbf{normalize $r_{a_{i,t}, t}$ to} $[0,1]$;
            \STATE $\hat{r}_{a,\,t} \gets 1 - \frac{{\bf 1}(a_{i,t}\,=\,a)}{p_{a,t}}\mleft(1\,-\,r_{a_{i,t}, t}\mright)$, $\forall a\in\calV_i$;
            \STATE $w_{a,t+1}\gets w_{a,t}\exp{\mleft(\eta_i^a \,\hat{r}_{a,t}\mright)}, \forall a\in\calV_i$;		
            \ENDFOR
	\end{algorithmic}\label{alg:action}
\end{algorithm}

We introduce the \actionsel algorithm and establish its performance guarantees. To this end, we first define the coordination problem that \actionsel addresses and relate it to  \Cref{pr:bandit} for each agent. 
\begin{itemize}[leftmargin=*]
    \item $\calA_t\triangleq \{a_{i,t}\}_{i\in\calN}$ is the set of all agents' actions at $t$;
    \item $\solopt\in\arg\max_{a_{i}\in\mathcal{V}_{i},\, \forall\, i\in\calN} f(\{a_{i}\}_{i\in \calN})$ is the optimal actions for agents $\calN$ that solve \cref{eq:intro-online}.
\end{itemize}

Intuitively, each agent $i$'s goal at each time step $t$ is to efficiently select an action $a_{i,t}$ that solves
\begin{equation}\label{eq:action-marginal-gain}
    \max_{a_{i,t}\in\calV_i} f(a_{i,t}\,|\,\{a_{j,t}\}_{j\in\calN_{i,t}}),
\end{equation}
To enable efficiency, the agents should ideally select actions \textit{simultaneously}, unlike offline algorithms such as Sequential Greedy~\cite{fisher1978analysis} and Resource-Aware distributed Greedy~\cite{xu2025communication} that employ (partially parallelized) sequential operations. But if they want to select actions simultaneously, $\{a_{j,t}\}_{j\in\calN_{i,t}}$ will become known only after agent $i$ selects $a_{i,t}$ and communicates with $\calN_{i,t}$. Therefore, computing the marginal gain is possible only in hindsight, once all agents' decisions have been finalized for time step $t$. Thus, action selection for each agent aligns with the framework of \Cref{pr:bandit}, where the reward of selecting $a_{i,t}\in\calV_i$ is $r_{a_{i,t},t}\triangleq f(a_{i,t}\,|\,\{a_{j,t}\}_{j\in\calN_{i,t}})$, which becomes available only after choosing $a_{i,t}$. 

\actionsel starts by initializing a learning rate $\eta_i^a$ and a weight vector $w_t$ for all available actions $a\in\calV_i$ (\Cref{alg:action}'s lines 1-2). Then, at each $t\in[T]$, it sequentially executes the following steps:
\begin{enumerate}[leftmargin=*]
    \item Compute probability distribution $p_t$ using $w_{t}$ (lines 3-4);
    \item Select action $a_{i,t}\in\calV_i$ by sampling from $p_t$ (line 5);
    \item Send $a_{i,t}$ to \neighborsel and receive neighbors' actions $\{a_{j,t}\}_{j\in\calN_{i,t}}$ (line 6);
    \item Compute reward $r_{a_{i,t},t}$, estimate reward $\hat{r}_{a,t}$ of each $a\in\calV_i$, and update weight $w_{a,t+1}$ of each $a\in\calV_i$ (lines 7-9).\footnote{{The coordination algorithms in~\cite{du2022jacobi,rezazadeh2023distributed,robey2021optimal} instruct the agents to select actions simultaneously at each time step as {\fontsize{7}{7}\selectfont\sc ActSel}, but they lift the coordination problem to the continuous domain and require each agent to know/estimate the gradient of the multilinear extension of $f$, which leads to a decision time at least one order higher than {\fontsize{7}{7}\selectfont\sc ActSel}~\cite{xu2025communication}.}}
\end{enumerate}

{We introduce the novel quantification that evaluates the benefit of information access in solving \Cref{pr:online}.
\begin{definition}[Value of Coordination ({\fontsize{9}{9}\selectfont\sf VoC})]\label{def:MI}
Consider $t\in [T]$ and agent $i\in\calN$ with action $a_{i,t}$ and coordination neighborhood $\calM_i$. Agent $i$'s \emph{Value of Coordination} is defined as: %
\begin{equation}\label{eq:SubMI}
    \smi{a_{i,t}}{\calN_{i,t}} \triangleq f(a_{i,t}) - f(a_{i,t}\,|\,\{a_{j,t}\}_{j\in\calN_{i,t}}),
\end{equation}where $\calN_{i,t}\subseteq\calM_i$ is the communication neighborhood that $i$ chooses to receive information from, $|\calN_{i,t}|\,\leq\alpha_i$.
\end{definition}
$\smi{a_{i,t}}{\calN_{i,t}}$ is thus an information theory metric similar to mutual information yet over the set function $f$. It represents the overlap between agent $i$ and its communication neighbors' actions. Particularly, $\sum_{t=1}^T\smi{a_{i,t}}{\calN_{i,t}}$ evaluates agent $i$'s benefit in coordinating with $\calN_{i,t}$ 
over the horizon $T$.
The larger is this value, the more related is the received information to the selected actions. 
}
\begin{definition}[Curvature~\cite{conforti1984submodular}]\label{def:curvature}
        The curvature of a normalized submodular function $f\colon 2^{\calV}\mapsto \mathbb{R}$ is defined as
        \begin{equation}
            \kappa_f\triangleq 1-\min_{\elem\in\calV}\frac{f(\calV)-f(\calV\setminus\{\elem\})}{f(\elem)}.
        \end{equation}
\end{definition} Intuitively, $\kappa_f$ measures how far~$f$ is from modularity: if $\kappa_f=0$, then  $f(\calV)-f(\calV\setminus\{v\})=f(v)$, $\forall v\in\calV$, \ie $f$ is modular. In~contrast, $\kappa_f=1$ in the extreme case where there exist $v\in\calV$ such that $f(\calV)=f(\calV\setminus\{v\})$, \ie $v$~has no contribution in the presence of $\calV\setminus\{v\}$.  
\begin{proposition}[Approximation Performance of \actionsel]\label{prop:action}
Over $t \in [T]$, agents $\calN$ can use \actionsel to select actions $\{\calA_t\}_{t\in[T]}$ such that
\begin{align}
    &\mathbb{E}\mleft[\frac{1}{T} \sum_{t=1}^T f(\calA_t)\mright] \geq (1-\curv_f) \, f(\solopt)\nonumber \\
    & + \sum_{i\in\calN} \mathbb{E}\mleft[\frac{1}{T} \sum_{t=1}^T\smi{a_{i,t}}{\calN_{i,t}}\mright] - \Tilde{O}\mleft(|\calN|\sqrt{\frac{|\bar{\calV}|}{T}}\mright). \label{eq:action_selection}
\end{align}where $|\bar{\calV}|\,\triangleq\max_{i\in\calN}|\calV_i|$.  %
\end{proposition}

\Cref{prop:action} implies that \actionsel's performance increases for neighborhoods $\{\calN_{i,t}\}_{i\in\calN}$ with higher $\smi{a_{i,t}}{\calN_{i,t}}$. The proof appears in Appendix~A.

\begin{lemma}[Monotonicity and Submodularity of {\fontsize{9}{9}\selectfont\sf VoC}]\label{lem:SubMI}
     Given an action $a\in\calV_i$ and a non-decreasing and 2nd-order submodular function $f\colon \calV_\calN \mapsto\mathbb{R}$, then $\smi{a}{\cdot}$ is non-decreasing and submodular in the second argument. 
\end{lemma}

\neighborsel will next leverage \Cref{lem:SubMI} to enable each agent to individually select its communication neighborhood that optimizes the approximation bound of \actionsel (\Cref{prop:action}).

\subsection{Neighbor Selection}\label{subsec:neighbor}

\setlength{\textfloatsep}{3mm}
\begin{algorithm}[t]
	\caption{\neighborsel for Agent $i$
	}
	\begin{algorithmic}[1]
		\REQUIRE \!Time horizon $T$; agent $i$'s coordination neighborhood $\calM_i$; agent $i$'s communication bandwidth $\alpha_i$. %
		\ENSURE \!Agent $i$'s communication neighborhood $\calN_{i,t}$, $\forall t\in[T]$.
		\medskip
  		\STATE $\eta_i^n\gets\sqrt{2\log{|\calM_i|}\,/\,{(|\calM_i|T)}}$; 
		\STATE $z_{1}^{(k)}\gets\mleft[z_{1,\,1}^{(k)}, \dots, z_{\alpha_i,1}^{(k)}\mright]^\top$ with $z_{j,1}^{(k)}=1$, $\forall v\in \calM_i, \forall k\in[\alpha_i]$;    
		\FOR {\text{each time step} $t\in [T]$}
            \STATE \textbf{receive} action $a_{i,t}$ \textbf{from} \actionsel;
            \FOR {$k = 1, \dots, \alpha_i$} 
            \STATE \textbf{get} distribution $q_{t}^{(k)}\gets z_t^{(k)}\,/\,{\|z_t^{(k)}\|_1}$; 
            \STATE \textbf{draw} agent $j_{t}^{(k)}\in\calM_i$ \textbf{from} $q_{t}^{(k)}$;
            \STATE \textbf{receive} action $a_{j_{t}^{(k)}, t}$ \textbf{from} $j_{t}^{(k)}$;
            \STATE $r_{j_{t}^{(k)}, t}\gets \smi{a_{i,t}}{\{a_{j_{t}^{(1)},t},\dots,a_{j_{t}^{(k)},t}\}} - $\\
            \hspace{1.28cm} $\smi{a_{i,t}}{\{a_{j_{t}^{(1)},t},\dots,a_{j_{t}^{(k-1)},t}\}}$ \\and
            \textbf{normalize $r_{j_{t}^{(k)}, t}$ to} $[0,1]$;
            \STATE $\hat{r}_{j,t}^{(k)} \gets 1 - \frac{{\bf 1}(j_{t}^{(k)}\,=\,j)}{q_{j,t}^{(k)}}\mleft(1\,-\,r_{j_{t}^{(k)},t}\mright)$, $\forall j\in\calM_i$;
            \STATE $z_{j,t+1}^{(k)}\gets z_{j,\,t}^{(k)}\exp{\mleft(\eta_i^n \,\hat{r}_{j,t}^{(k)}\mright)}$, $\forall j\in\calM_i$;					\ENDFOR
                \STATE $\calN_{i,t}\gets\{j_{k, t}\}_{k\in [\alpha_i]}$;
		\ENDFOR
	\end{algorithmic}\label{alg:neighbor}
\end{algorithm}

We now present the \neighborsel algorithm and its performance guarantees for maximizing {\fontsize{9}{9}\selectfont\sf VoC}. Particularly, we introduce the neighbor-selection problem that \neighborsel enables agents to solve in parallel and demonstrate that it aligns with~\Cref{pr:bandit}.

Since the suboptimality bound of \actionsel in~\cref{eq:action_selection} improves as $\smi{a_{i,t}}{\calN_{i,t}}$ increases, \neighborsel aims to enable each agent $i$ to choose neighbors by solving the following cardinality-constrained maximization problem:
{\begin{equation}\label{eq:neighbor_selection}
    \max_{\calN_{i,t}\,\subseteq\,\calM_i,\,|\calN_{i,t}|\,\leq\,\alpha_i} \; \sum_{t=1}^{T} \;\smi{a_{i,t}}{\calN_{i,t}},
\end{equation}}where $a_{i,t}$ is given by \actionsel. {This is a submodular maximization problem since we show in~\Cref{lem:SubMI} that $ \smi{a_{i,t}}{\calN_{i,t}}$ is submodular in $\calN_{i,t}$. But $\smi{a_{i,t}}{\calN_{i,t}}$ is computable in hindsight only: $\{a_{j,t}\}_{j\in\calN_{i,t}}$ becomes known only after agent $i$ has selected and communicated with $\calN_{i,t}$. Therefore, \cref{eq:neighbor_selection} takes the form of cardinality-constrained bandit submodular maximization~\cite{zhang2019online,matsuoka2021tracking,xu2023bandit}, which is an extension of \Cref{pr:bandit} to the multi-agent submodular setting.

Solving \cref{eq:neighbor_selection} using algorithms for \Cref{pr:bandit} will lead to exponential-running-time algorithms due to an exponentially large $\calV$ per \cref{eq:bandit}~\cite{matsuoka2021tracking}. Therefore, \neighborsel instead extends \cite[Algorithm 2]{matsuoka2021tracking}, which can solve \cref{eq:neighbor_selection} in the full-information setting, to the bandit setting~\cite{xu2023bandit}. Specifically, \neighborsel decomposes \cref{eq:neighbor_selection} to $\alpha_i$ instances of \Cref{pr:bandit}, and separately solves each of them using \expthree.

\neighborsel starts by initializing a learning rate $\eta_i^n$ and $\alpha_i$ weight vectors $z_t^{(k)}, \forall k\in[\alpha_i]$, each determining the $k$-th selection in $\calN_{i,t}$ (\Cref{alg:neighbor}'s lines 1-2). Then, at each $t\in[T]$, it sequentially executes the following steps:
\begin{enumerate}[leftmargin=*]
    \item Receive action $a_{i,t}$ by \actionsel (lines 3-4);
    \item Compute distribution $q_t^{(k)}$ using $z_{t}^{(k)}, \forall k\in [\alpha_i]$ (lines 5-6);
    \item Select agent $j_{t}^{(k)}\in\calM_i$ as neighbor by sampling from $q_t^{(k)}$, and receive its action $a_{j_{t}^{(k)},t}, \forall k\in [\alpha_i]$ (lines 7-8);
    \item For each $k\in[\alpha_i]$, compute reward $r_{j_{t}^{(k)},t}$ associated with each $j_{t}^{(k)}$, estimate reward $\tilde{r}_{j,t}^{(k)}$ of each $j\in\calM_i$,
    and update weight $z_{j,t+1}^{(k)}$ of each $j\in\calM_i$ (lines 9-12).
\end{enumerate}

\section{Suboptimality Guarantees}\label{sec:suboptimality-guarantees}
We show that \alg's approximation performance against the optimal solution to \cref{eq:intro-online}  improves with the sum of all agents' {\fontsize{9}{9}\selectfont\sf VoC}, and is strictly positive anytime, even before convergence.
To this end, we begin with an a priori bound 
validating that \neighborsel indeed improves \alg's performance bound (\Cref{subsec:priori-bound}). We then provide an anytime strictly positive a posteriori bound (\Cref{subsec:posteriori-bound}). 
Combining the first two results, we finally present an asymptotic bound (\Cref{subsec:conservative-bound}). 
All proofs appear in Appendix~A.

\subsection{A Priori Bound}\label{subsec:priori-bound}
To present the a priori bound,  we first present the following definitions and lemmas that measure the performances of \actionsel and \neighborsel. We use the following notation:
\begin{itemize}[leftmargin=*]
    \item $\kappa_{I,i} \triangleq \max_{t\in[T]} \kappa_{I,i,t}$, where $\kappa_{I,i,t}$ is the curvature of $\smi{a_{i,t}}{\cdot}$, $\forall t\in[T]$. $\kappa_{I,i}$ is independent of $\kappa_f$;
    \item $\rho(\kappa, \alpha) \triangleq \kappa^{-1}\mleft[1-(1-\kappa/\alpha)^{\alpha}\mright]$, where $\kappa\in[0,1]$ and $\alpha\in\mathbb{N}$. Notice that $1 \geq \kappa^{-1}\mleft[1-(1-\kappa/\alpha)^{\alpha}\mright] \stackrel{\alpha\rightarrow\infty}{>} \kappa^{-1}\mleft(1-e^{-\kappa}\mright) \stackrel{\kappa\rightarrow 1}{\geq} 1-e^{-1}$.
\end{itemize}

The performances of \actionsel and \neighborsel are measured by the following quantities:
\begin{definition}[Static Regret of Action Selection]\label{def:action-regret}
    Given agent $i$ has neighbors $\{\calN_{i,t}\}_{t\in[T]}$ over the horizon $[T]$. At each time step $t$, suppose agent $i$ selects an action $a_{i,t}$. Then, the static regret of $\{a_{i,t}\}_{t\in [T]}$ is defined as
    \begin{align}\label{eq:action}
        &\operatorname{A-Reg}_T\mleft(\{a_{i,t}\}_{t\in [T]}\mright) \triangleq \\\nonumber
        &\max_{a\in\calV_i} \sum_{t=1}^{T} f(a\,|\,\{a_{j,t}\}_{j\in\calN_{i,t}}) - \sum_{t=1}^{T} f(a_{i,t}\,|\,\{a_{j,t}\}_{j\in\calN_{i,t}}).
    \end{align}
\end{definition}
\begin{definition}[$\rho(\kappa_{I,i}, \alpha_i)$-Approximate Static Regret of Neighbor Selection]\label{def:NReg}
    Given agent $i$ has actions $\{a_{i,t}\}_{t\in[T]}$ over horizon $[T]$. At each time step $t$, suppose agent $i$ selects a set of neighbors $\calN_{i,t}\subseteq\calM_i$, $|\calN_{i,t}|\,\leq\alpha_i$. Then, the $\rho(\kappa_{I,i}, \alpha_i)$-approximate static regret of $\{\calN_{i,t}\}_{t\myin[T]}$ is defined as
    \begin{align}%
        &\operatorname{N-Reg}_{\{a_{i,t}\}_{t\in [T]}}^{\rho(\kappa_{I,i}, \alpha_i)}\mleft(\{\calN_{i,t}\}_{t\in [T]}\mright) \nonumber \\
        &\triangleq \rho(\kappa_{I,i}, \alpha_i) \max_{\mathcal{N}_{i,t} \subseteq \mathcal{M}_i, |\mathcal{N}_{i,t}| \leq \alpha_i}\sum_{t=1}^{T} \smi{a_{i,t}}{\calN_{i,t}} \nonumber\\\label{eq:regret_1}
        &\quad- \sum_{t=1}^{T}\smi{a_{i,t}}{\calN_{i,t}}.
    \end{align}
\end{definition}\Cref{def:NReg} evaluates the suboptimality of $\{\calN_{i,t}\}_{t\in [T]}$ against the optimal communication neighborhood that would have been selected if $\smi{a_{i,t}}{\cdot}$ had been known a priori, $\forall t\in [T]$. The optimal value in \cref{eq:regret_1} is discounted by $\rho(\kappa_{I,i}, \alpha_i)$ since the problem in \cref{eq:neighbor_selection} is NP-hard to solve with an approximation factor greater than $\rho(\kappa_{I,i}, \alpha_i)$ even when $\smi{a_{i,t}}{\cdot}$ is known a priori~\cite{conforti1984submodular}.

\begin{lemma}[Suboptimality Guarantee of \actionsel]\label{lem:AReg-bound}
    Consider horizon $[T]$. If agent $i$ has a sequence of neighbors $\{\calN_{i,t}\}_{t\in [T]}$, regardless of how they are selected, then using \actionsel, agent $i$ can choose actions $\{a_{i,t}\}_{t\in [T]}$ such that
    \begin{equation}
        \mathbb{E}\mleft[\operatorname{A-Reg}_T\mleft(\{a_{i,t}\}_{t\in [T]}\mright)\mright] \leq \Tilde{O}\mleft(\sqrt{|\calV_i|\,T}\mright),
    \end{equation}given the learning rate is set as $\eta_i^a=\sqrt{2\log{|\calV_i|}\,/\,(|\calV_i|\,T)}$.
\end{lemma}

\begin{lemma}[Suboptimality Guarantee of \neighborsel]\label{lem:NReg-bound}
    Consider horizon $[T]$. If agent $i$ has a sequence of actions $\{a_{i,t}\}_{t\in [T]}$, regardless of how they are selected, then using \neighborsel, agent $i$ can choose neighborhoods $\{\calN_{i,t}\}_{t\in [T]}$ such that
    \begin{equation}
        \mathbb{E}\mleft[\operatorname{N-Reg}_{\{a_{i,t}\}_{t\in [T]}}^{\rho(\kappa_{I,i}, \alpha_i)}\mleft(\{\calN_{i,t}\}_{t\in [T]}\mright)\mright] \leq \Tilde{O}\mleft(\alpha_i\sqrt{|\calM_i|\,T}\mright),
    \end{equation}given the learning rate set as $\eta_i^n=\sqrt{2\log{|\calM_i|}\,/\,(|\calV_i|\,T)}$.
\end{lemma}

Now we present the a priori bound of \alg.

\begin{theorem}[A Priori Approximation Performance]\label{th:main}Using \alg, each agent $i$  selects actions $\{a_{i,t}\}_{t\in[T]}$ and communication neighborhoods $\{\calN_{i,t}\}_{t\in[T]}$ that guarantee:
\begin{align}
    &\mathbb{E}\mleft[\frac{1}{T} \sum_{t=1}^{T}f(\calA_t)\mright] \geq (1-\curv_f)\, f(\solopt) \nonumber\\
    &+ \kappa_f\, (1-\kappa_f)\, \rho(\kappa_{I}, \bar{\alpha}) \nonumber \\
    &\hspace{.7cm}\times \sum_{i\in\calN} \mathbb{E}\mleft[\frac{1}{T} \sum_{t=1}^{T}\smi{a_{i,t}}{\calN_{i}^{\star}(\{a_{i,t}\}_{t\in [T]}; \alpha_i, \calM_i)}\mright] \nonumber\\
    &- \Tilde{O}\mleft(|\calN|\sqrt{{\mleft(\bar{\alpha}^2\,|\bar{\calM}|\,+\,|\bar{\calV}|\mright)}/{T}}\mright),\label{eq:thm-3}
\end{align}where $\kappa_{I}\triangleq\max_{i\in\calN}\kappa_{I,i}$, $\bar{\alpha}\triangleq\max_{i\in\calN}\alpha_i$, $|\bar{\calV}|\,\triangleq\max_{i\in\calN}|\calV_i|$, $|\bar{\calM}|\,\triangleq\max_{i\in\calN}|\calM_i|$, $\kappa_f, \kappa_I \in[0,1]$, $\rho(\kappa_{I}, \bar{\alpha})\in[1-1/e, 1]$, and $\calN_{i}^{\star}(\{a_{i,t}\}_{t\in [T]}; \alpha_i, \calM_i)\subseteq\calM_i$ is the optimal communication neighborhood that solves~\cref{eq:neighbor_selection} given $\{a_{i,t}\}_{t\in [T]}$ subject to $\alpha_i$. 
The expectation is due to the algorithm's internal randomness. %
\end{theorem}

{\Cref{th:main} implies that as $T\rightarrow\infty$, the a priori approximation ratio spans the interval between fully centralized and fully decentralized coordination in accordance to \voc, and the spectrum is depicted in red in Fig.~\ref{fig:asymp-approx-ratio}. In particular, when the network is fully centralized with all agents listening to all others, \ie $\calN_{i,t}\equiv\calN\setminus\{i\}$,
    \begin{equation}\label{eq:thm-1}
    \hspace{-1.5mm}\mathbb{E}\mleft[\frac{1}{T} \sum_{t=1}^{T} f(\calA_t)\mright] \geq \frac{1}{1+\kappa_f}\,f(\solopt) - \Tilde{O}\mleft(|\calN|\sqrt{\frac{|\bar{\calV}|}{T}}\mright),
\end{equation}where the guaranteed $1/(1+\kappa_f)$ bound is near-optimal~\cite{conforti1984submodular}. When the network is fully decentralized with no agent listening to any others, \ie $\calN_{i,t}\equiv\emptyset$,
\begin{equation}\label{eq:thm-2}
    \hspace{-1.5mm}\mathbb{E}\mleft[\frac{1}{T} \sum_{t=1}^{T}f(\calA_t)\mright] \geq (1-\curv_f) \,f(\solopt) - \Tilde{O}\mleft(|\calN|\sqrt{\frac{|\bar{\calV}|}{T}}\mright),
\end{equation}
where $1-\kappa_f$ is the worst-case bound achieved by \alg.}

\begin{figure}[t]
    \captionsetup{font=footnotesize}
    \centering
    \includegraphics[width=\columnwidth]{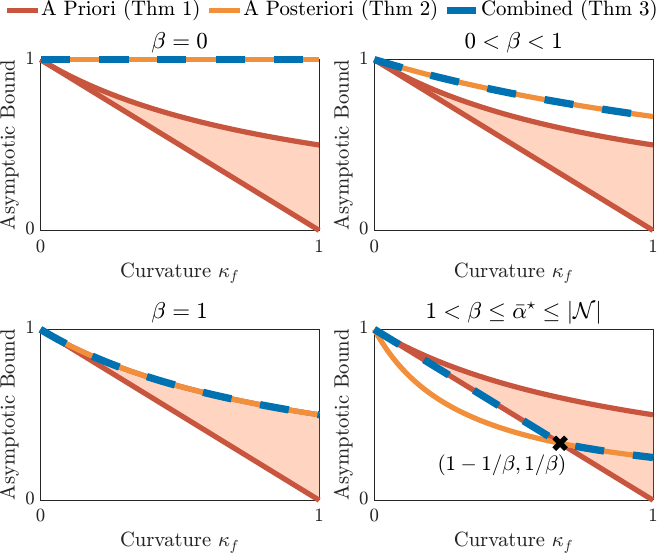}
    \vspace{-3mm}
    \caption{\textbf{Asymptotic approximation bounds of {\fontsize{8}{8}\selectfont\sc Anaconda}.} As $T\rightarrow\infty$, the bounds provided by \Cref{th:main,th:posteriori,th:asymptotic} are shown with varying ranges of $\kappa_f$ and achieved $\beta$ (defined in~\cref{eq:beta}). The a priori bound (red) varies with the sum of all agents' {\fontsize{7}{7}\selectfont\sf VoC}; the a posteriori bound (orange) decreases as $\beta$ increases; and the combined bound (blue) takes the maximum of the a priori lower bound and the a posteriori bound. 
    }\label{fig:asymp-approx-ratio}
   \vspace{-3mm}
\end{figure}

\subsection{A Posteriori Bound}\label{subsec:posteriori-bound}

We provide an a posteriori bound for \alg that is strictly positive, even for finite $T$.

\begin{theorem}[A Posteriori Approximation Performance]\label{th:posteriori} 
Using \alg, each agent $i\in\calN$  selects $\{a_{i,t}\}_{t\in[T]}$ and $\{\calN_{i,t}\}_{t\in[T]}$ that guarantee:

\begin{equation}
    \mathbb{E}\mleft[\frac{1}{T}\sum_{t=1}^T f(\calA_t)\mright] \geq \frac{1}{1 + \beta \kappa_f + \Tilde{O}\mleft(|\calN|\sqrt{|\bar{\calV}|/T}\mright)} \, f(\solopt), \label{eq:thm-posteriori}
\end{equation}where 
\begin{equation}\label{eq:beta}
    \beta\triangleq\frac{\sum_{t=1}^{T}\sum_{i\in\calN}f(a_{i,t}\,|\,\{a_{j,t}\}_{j\in\calN_{i,t}})}{\sum_{t=1}^{T}f(\calA_t)} \in [0, \bar\alpha^\star]
\end{equation}is computable after \alg terminates. Here, $\alpha^\star(\cdot)$ denotes the fractional independence number of a graph~\cite{godsil2013algebraic}. For each $t\in[T]$, let
$\calG_t^{\mathsf{DAG}}=(\calN,\calE_t^{\mathsf{DAG}})$, with
$\calE_t^{\mathsf{DAG}}\subseteq \calE_t$, be an acyclic spanning subgraph of
$\calG_t$ satisfying
\begin{equation}
    \calG_t^{\mathsf{DAG}}
    \in
    \arg\min_{
    \calH\,=\,(\calN,\, \calF):\, \calF\, \subseteq\, \calE_t}
    \alpha^\star(\calH)\;\; \text{s.t.}\;\; \calH\text{ is a DAG}.
\end{equation}
Define
\begin{equation}
    \bar\alpha^\star
    \triangleq
    \max_{t\in[T]}\alpha^\star(\calG_t^{\mathsf{DAG}}).
\end{equation}
Then $\bar\alpha^\star\le |\calN|$. The expectations are taken with respect to
the internal randomness of \alg.
\end{theorem}

The bound is depicted in Fig.~\ref{fig:asymp-approx-ratio} in orange for varying ranges of $\beta$. Since $\beta \le \bar\alpha^\star$, \Cref{th:posteriori} implies that \alg always achieves a strictly positive asymptotic approximation ratio:
\begin{align}\label{eq:asymp-posteriori}
        \mathbb{E}\mleft[\frac{1}{T} \sum_{t=1}^T f(\calA_t)\mright] &\stackrel{T\rightarrow\infty}{\geq} \frac{1}{1 + \kappa_f\beta}\, f(\solopt) \nonumber \\
        &\hspace{2.5mm}\geq\frac{1}{1 + \kappa_f\bar\alpha^\star}\, f(\solopt) > 0.
\end{align}Moreover, if $\beta=0$, \ie each agent's selected action fully overlaps with its neighbors' selected actions, then \alg is asymptotically optimal:
\begin{equation}
    \mathbb{E}\mleft[\frac{1}{T} \sum_{t=1}^T f(\calA_t)\mright] \stackrel{T\rightarrow\infty}{\geq} f(\solopt).
\end{equation}Finally, if $0\leq\beta\leq1$, then \alg asymptotically outperforms the bound of Sequential Greedy~\cite{fisher1978analysis}:
\begin{equation}
    \mathbb{E}\mleft[\frac{1}{T} \sum_{t=1}^T f(\calA_t)\mright] \stackrel{T\rightarrow\infty}{\geq} \frac{1}{1+\kappa_f}\, f(\solopt).
\end{equation}
This last scenario is guaranteed to occur when, \eg there exists an order of the agents such that $[i-1] \subseteq \calN_{i,t}, \forall i, \forall t$.

\subsection{Combined Asymptotic Bound}\label{subsec:conservative-bound}

Combining \Cref{th:main,th:posteriori}, we get: 

\begin{theorem}[Asymptotic Approximation Performance]\label{th:asymptotic} 
\alg asymptotically achieves:
\begin{equation}
    \hspace{-2.45mm}\frac{\mathbb{E}\mleft[\frac{1}{T}\sum_{t=1}^T f(\calA_t)\mright]}{f(\solopt)} \stackrel{T\rightarrow\infty}{\geq} \max{\mleft(1-\kappa_f, \frac{1}{1 + \beta\kappa_f}\mright)} > 0, \label{eq:thm-asymptotic}
\end{equation}where $\kappa_f\in [0,1]$ and $\beta\in [0,\bar\alpha^\star]$. %
\end{theorem}
The bound is depicted in Fig.~\ref{fig:asymp-approx-ratio} in blue.

\section{Decision Time Analysis}\label{sec:resource-guarantees}
We present the convergence time of \alg by analyzing its computation and communication complexity, accounting for the delays due to function evaluation and the transmission of messages under realistic communications. 
\begin{itemize}[leftmargin=*]
    \item $\tau_f$ is the time required for one evaluation of $f$;
    \item $\tau_c$ is the time for agent $i$ to transmit the information about one action to agent $j$, without loss of generality, for all $(i\rightarrow j)\in\calE_t, \forall t$.
\end{itemize}
\begin{theorem}[Convergence Time]\label{th:speed}
\alg converges in $O\mleft[(\tau_f\,\bar{\alpha}\,+\,\tau_c\,)(\bar{\alpha}^2\,|\bar{\calM}|\,+\,|\bar{\calV}|)|\calN|^2\,/\,\epsilon\mright]$ time.
\end{theorem}
In sparse networks, the following corollary holds:
\begin{corollary}[Convergence Time for Sparse Networks]\label{cor:speed}
In sparse networks, where $|\calM_i|\,=o(|\calN|), \forall i\in\calN$, \alg converges in $O\mleft[(\tau_f\,\bar{\alpha}\,+\,\tau_c\,)|\bar{\calV}|\,|\calN|^2\,/\,\epsilon\mright]$ time.
\end{corollary}
The proof appears in Appendix~B.

\section{Necessity for Network Optimization: Numerical Evaluation in Area Monitoring}\label{sec:experiments-1}
\begin{figure}[t]
    \captionsetup{font=footnotesize}
    \centering
    \includegraphics[width=\columnwidth]{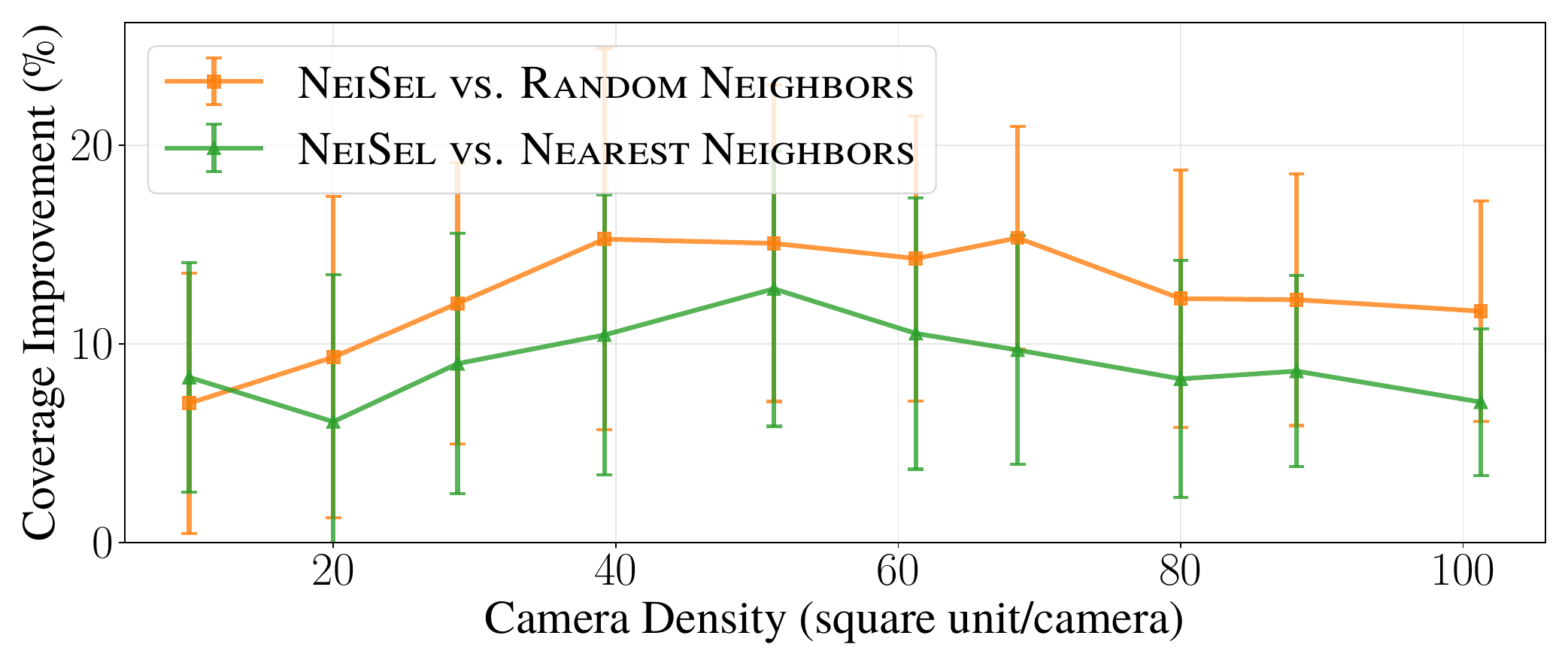}
    \vspace{-5mm}
    \caption{\textbf{Comparison of neighbor selection strategies with varying network density.} Across 20 MC trials each with 2000 decision rounds, we compare {\footnotesize \sc NeiSel} with two benchmark strategies, Nearest Neighbors and Random Neighbors. We tune the network density by varying the map area while fixing the network size at 20 agents: as the camera density grows, the network becomes sparser. 
    }\label{fig:comparison-neighbor-selection}
   \vspace{-3mm}
\end{figure}

In this section, we demonstrate that optimizing the network topology, per \alg, leads to improved coordination performance compared to heuristics for network design, such as the nearest and random selection that are typically used in controls and robotics~\cite{jadbabaie2003coordination,zhou2023racer,liu2025slideslam,xu2025communication}. 
In particular, we compare the proposed \neighborsel (\Cref{alg:neighbor}) with two heuristic strategies, \textit{Nearest Neighbors} and \textit{Random Neighbors}, in simulated 2D area-monitoring scenarios. The results show that \neighborsel consistently outperforms the benchmarks across different network densities (Fig.~\ref{fig:comparison-neighbor-selection}), and that while nearest-neighbor heuristic is quite \textit{misleading} in certain structured environments, \neighborsel can configure the optimal communication neighborhood (Fig.~\ref{fig:urban}). We defer the description of the former simulations to Appendix~V and describe the latter below.

In more detail, 
eight cameras are deployed to monitor four street blocks as shown in Fig.~\ref{fig:urban}. Under $\alpha_i = 1$ for all $i \in \calN$, the optimal neighbor for each camera is the one positioned opposite its corresponding street block so as to minimize FOV overlap. 
The results, shown in Fig.~\ref{fig:urban} and \Cref{tab:network-comparison}, indicate that \neighborsel outperforms the two heuristic baselines by $27.2\%$ and $11.5\%$ in mean coverage, respectively. We next describe the simulation setup, compared algorithms, and results.

\newcommand{\introFigTitleWidth}{0.2cm}
\newcommand{\introFigColWidth}{5.7cm}
\newcommand{\introFigSpacing}{\hspace{-2mm}}
\newcommand{\intoFigNameSpacing}{}
\newcommand{\advFigColWidth}{6cm}

\begin{figure*}[t]
    \captionsetup{font=footnotesize}
	\begin{center}
	\hspace{-9.8cm}
      \begin{minipage}{\columnwidth}
            \begin{tabular}{p{\introFigColWidth}p{\introFigColWidth}p{\introFigColWidth}}

            \begin{minipage}{\introFigColWidth}%
                  \centering%
                  \includegraphics[width=\columnwidth]{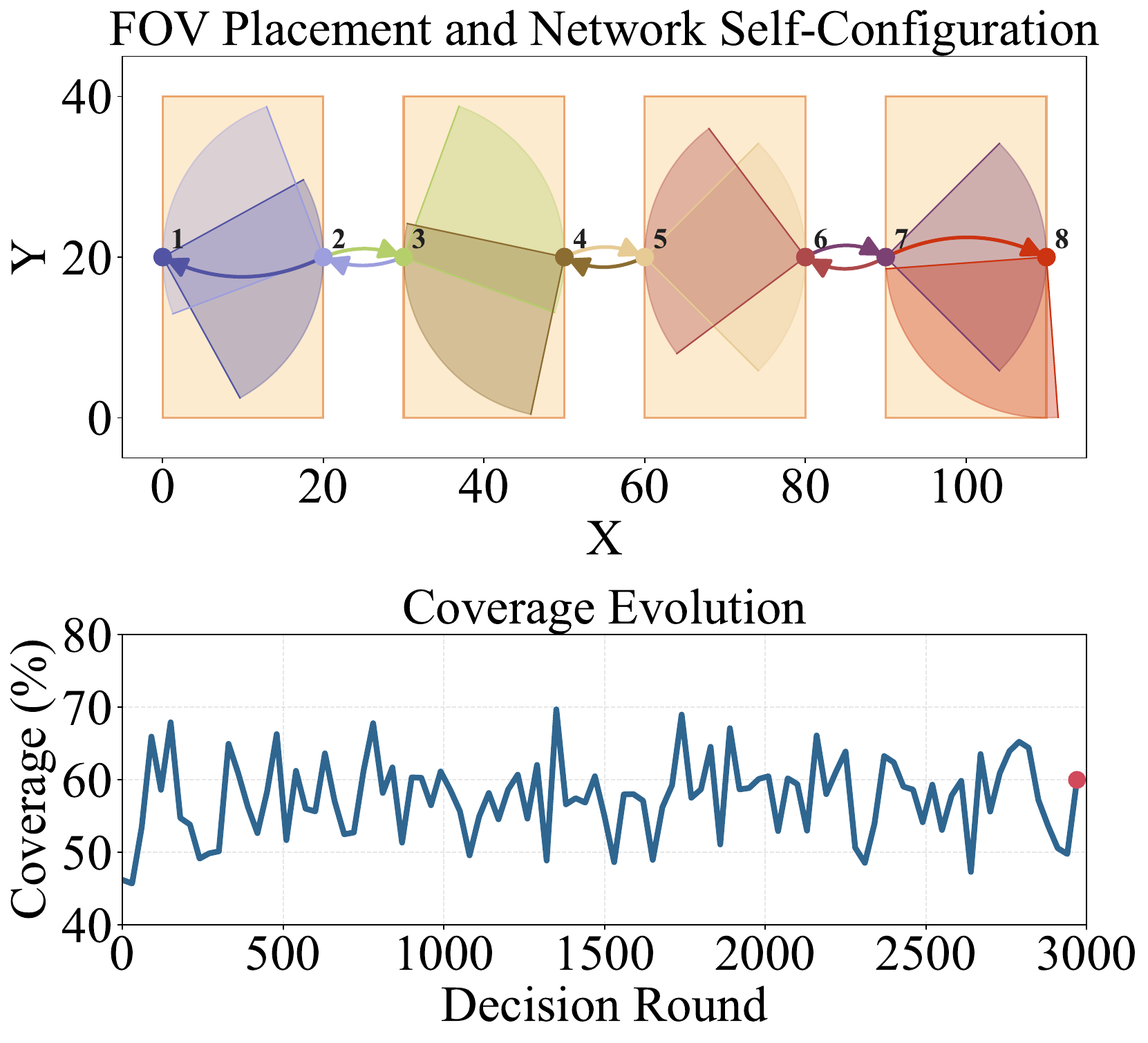} \\
                  \caption*{(a) {\fontsize{8}{8}\selectfont\sc ActSel} + Nearest neighbors.
                  }
            \end{minipage}
            &
            \begin{minipage}{\introFigColWidth}%
                  \centering%
                  \includegraphics[width=\columnwidth]{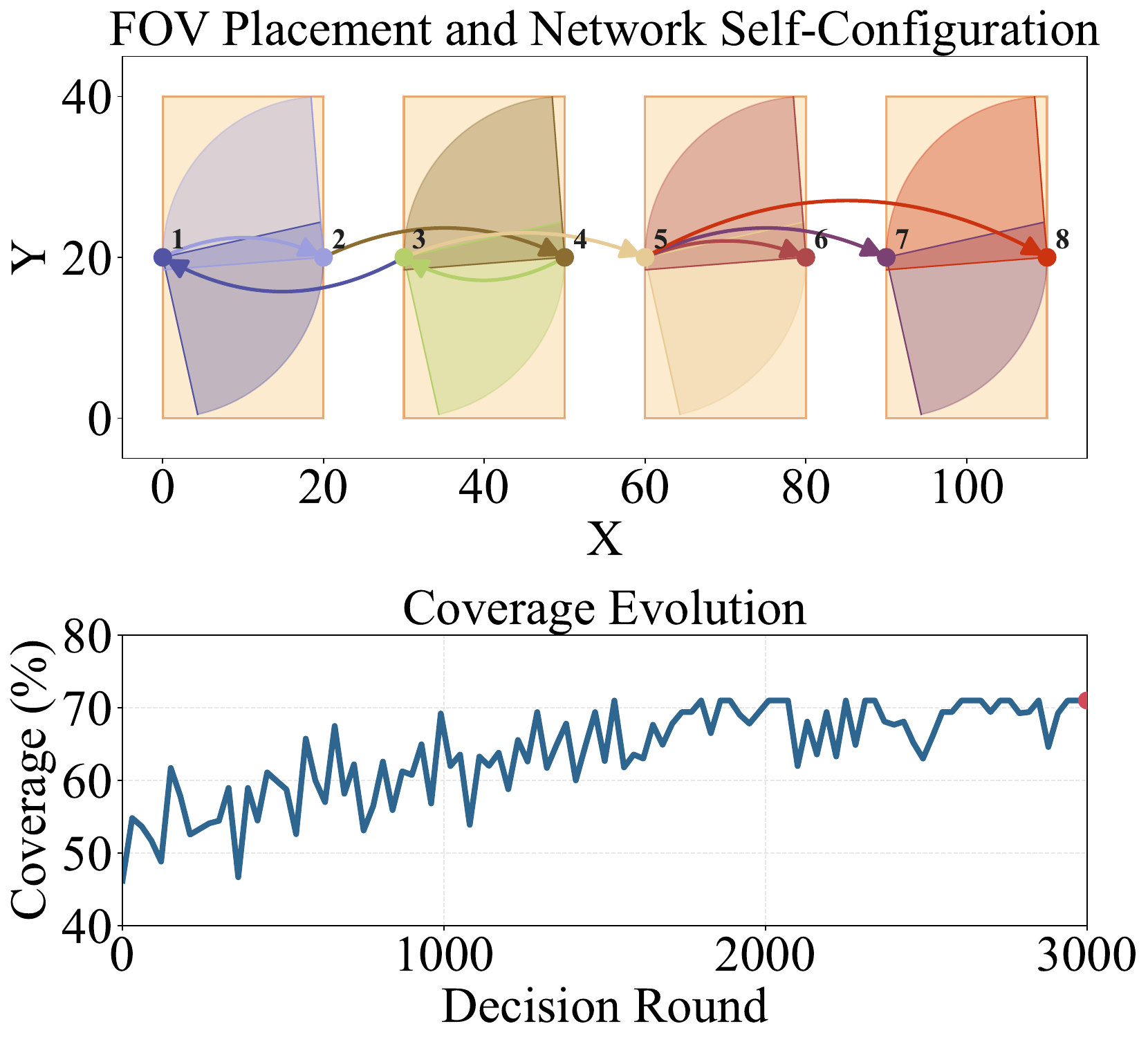} \\ 
                  \caption*{(b) {\fontsize{8}{8}\selectfont\sc ActSel} + Random neighbors.
                  }
            \end{minipage}
            &
            \begin{minipage}{\introFigColWidth}%
                  \centering%
                  \includegraphics[width=\columnwidth]{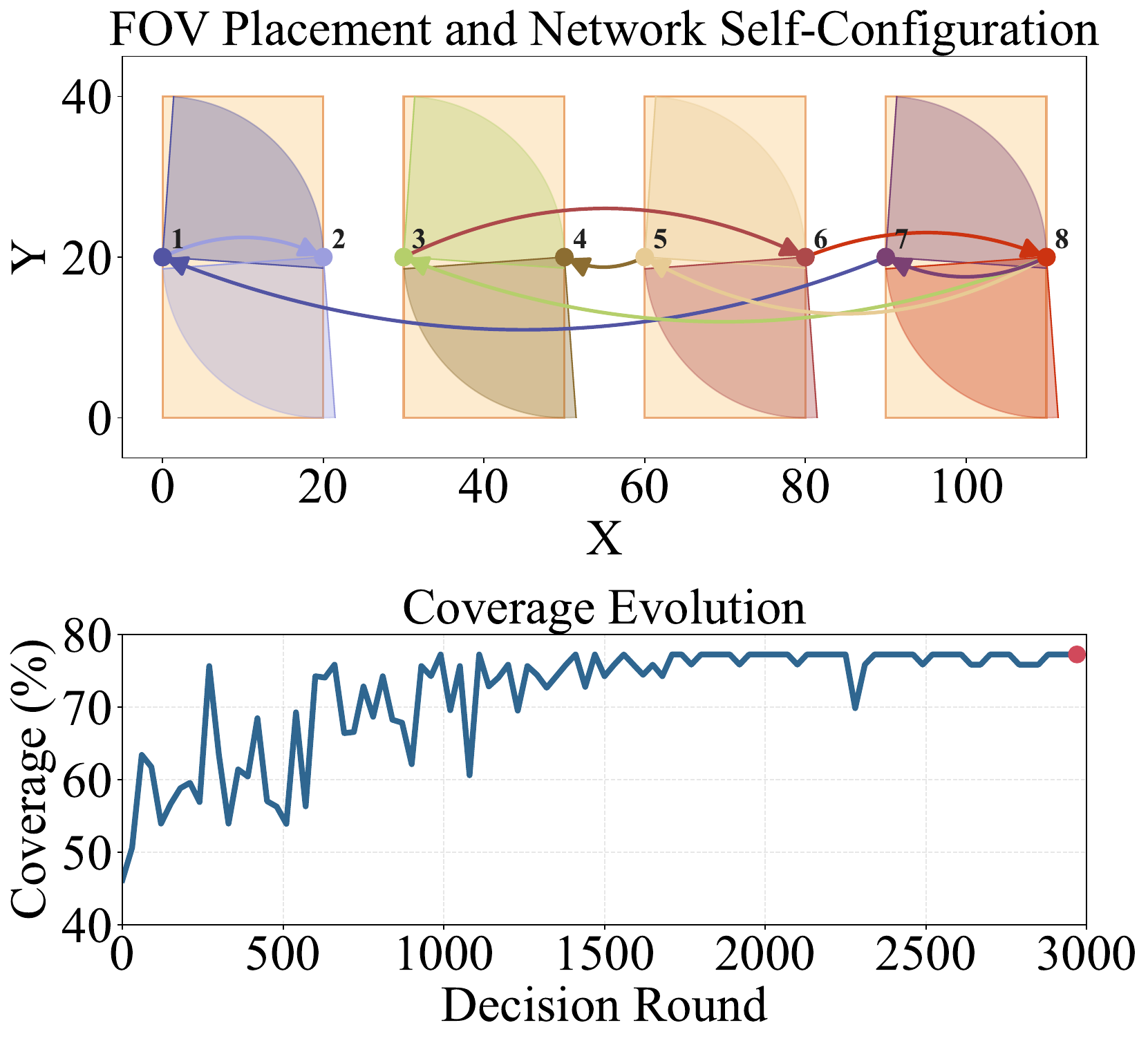} \\
                  \caption*{
                  (c) {\fontsize{8}{8}\selectfont\sc ActSel + NeiSel (Anaconda)}.
                  }
            \end{minipage}
            \end{tabular}
	\end{minipage} 
	\caption{\textbf{Comparison of neighbor selection strategies in a structured environment.} Three algorithms are compared with the same action selection strategy {\fontsize{8}{8}\selectfont\sc ActSel} but different neighbor selection strategies ({\fontsize{8}{8}\selectfont\sc NeiSel} vs. nearest neighbors vs. random neighbors) in the same structured environment. %
	}\label{fig:urban}
	\vspace{-7mm}
	\end{center}
\end{figure*}

\myParagraph{Setup} 
    \textit{Environment:} The environment is a $110\times 40$ map with four $20\times 40$ street blocks to monitor, as in Fig.~\ref{fig:urban}. 
    \textit{Agents:} Eight cameras are located on the boundaries of street blocks with locations $(0, 20)$, $(20, 20)$, $(30, 20)$, $(50, 20)$, $(60, 20)$, $(80, 20)$, $(90, 20)$, and $(110, 20)$. They all have large enough communication ranges $c_i$ such that $\calM_i=\calN\setminus\{i\}, \forall i\in\calN$.
    \textit{Actions:} All cameras $i\in\calN$ have FOV radius $r_i=20$ and AOV $\theta_i=\pi/2$, with direction $a_{i,t}$ chosen from the 16 cardinal directions $\calV_i$, $\forall t$. Each camera $i$ is considered unaware of $\calV_j, j\in\calN\setminus\{i\}$. Thus, the cameras have to communicate to know about one another's action information. 
    \textit{Communication Network:} The emergent time-varying communication network $\calG_t$ can be directed and disconnected. At each decision round $t$, each camera $i$ will first find its coordination neighborhood $\calM_i\triangleq\{j\}_{\|x_j-x_i\|\leq c_i,\, j\myin\calN\setminus\{i\}}$, where $c_i$ is $i$'s communication range. Then, it will select communication neighborhood $\calN_{i,t}$ from $\calM_i$, following a strategy that will be determined by different compared algorithms. Once $\calN_{i,t}$ is configured by all $i\in\calN$, then $\calG_t$ is defined. 
    \textit{Objective Function:}  $f(\{a_{i,t}\}_{i\in \calN})$ is the total area of interest covered by the cameras $\calN$ when they select $\{a_{i,t}\}_{i\in \calN}$ as their FOV directions. $f$ is proved to be submodular~\cite{corah2018distributed}.

\begin{table}[tbp]
\captionsetup{font=footnotesize}
\centering
\caption{Comparison of coverage performance (\%) in the urban scenario as in Fig.~\ref{fig:urban} with different neighbor selection strategies. The best coverage performance is in \textbf{bold}.}
\setlength{\tabcolsep}{4pt}
\resizebox{\columnwidth}{!}{%
\begin{tabular}{@{}l|ccc@{}}
\toprule
\textbf{Metric} & \textbf{Nearest Neighbors} & \textbf{Random Neighbors} & \textbf{Ours} \\
\midrule
Mean $\pm$ Std (\%)  & 56.53 $\pm$ 4.57 & 64.47 $\pm$ 6.07 & \textbf{71.89 $\pm$ 7.36} \\
Min (\%)   & 35.56 ($t{=}1$) & 37.94 ($t{=}113$)  & \textbf{39.72} ($t{=}169$) \\
Max (\%)   & 71.09 ($t{=}2484$) & 72.56 ($t{=}895$)  & \textbf{77.25} ($t{=}619$) \\
\bottomrule
\end{tabular}%
}
\label{tab:network-comparison}
\end{table}

\myParagraph{Performance Metrics} We evaluate the achieved coverage of street blocks of the algorithms over 3000 decision rounds.

\myParagraph{Compared Algorithms}
We evaluate the following benchmarks, all selecting one neighbor per camera at each decision round, \ie $\alpha_i=1$ for all $i\in\calN$:
    \textit{(i) \alg:} Our algorithm will follow the process described in \Cref{sec:algorithm}. That is, at each round, the agents will each first sample its action $a_{i,t}\in\calV_i$ per \actionsel and neighbors $\calN_{i,t}\subseteq\calM_i$ per \neighborsel. Then, they will each listen to $\calN_{i,t}$ and use the received information to update the two bandit algorithms.
    \textit{(ii) \actionsel + Nearest Neighbors:} Action selection follows \alg, while each camera $i$ will always have the same closest agent $j\in\calM_i$ as its neighbor. %
    \textit{(iii) \actionsel + Random Neighbors:} Action selection follows \alg, while each camera $i$ will uniformly resample its neighbor from $\calM_i$ at each decision round. 

\myParagraph{Results} {We observe that network optimization via \neighborsel increases both coverage performance and convergence speed, as shown in Fig.~\ref{fig:urban}.} In particular, \alg outperforms the two benchmarks in all aspects of mean, minimum, and maximum coverage. It improves the benchmarks by $(71.89\%/56.53\%-1=)27.2\%$ and $(71.89\%/64.47\%-1=)11.5\%$ in the mean coverage, respectively (\Cref{tab:network-comparison}). Moreover, \neighborsel converges to the optimal network configuration after $\sim$1000 rounds, and then \actionsel also converges to the optimal coverage performance (77.25\%) after $\sim$1800 rounds (Fig.~\ref{fig:urban}(c)). In contrast, we do not observe any convergence for the nearest neighbor selection (Fig.~\ref{fig:urban}(a)), and the convergence appears slower for random neighbor selection after $\sim$2400 rounds, with a suboptimal coverage performance (72.56\%). 

The reason why random neighbor selection performs better than nearest neighbor selection is that the latter will always select the same neighbors, thus never choose the optimal configuration. In contrast, random selection can happen to pick the optimum by chance, thus providing better performance.

\section{Numerical Evaluation in Area Monitoring}\label{sec:experiments-2}

While much of the distributed optimization literature analyzes computation and communication complexities, it rarely considers how the resulting delays influence the algorithm’s practical optimality over time. 
In time-critical applications, however, the ability to make rapid decisions is essential, and computation and communication delays directly decide how many coordination rounds can be completed within a finite time horizon. As a result, an algorithm with a lower theoretical bound but higher decision frequency may outperform another with a higher bound but lower frequency, and thus delay-aware evaluation is crucial for revealing such effects. 

{To this end, in the section, we demonstrate the benefits of \alg by comparing it with benchmarks under two settings, \ie with and without communication and computation delays. In particular, \alg outperforms because it runs fast under delays.}
The simulation results, summarized in Figs.~\ref{fig:comparison-no-delay}--\ref{fig:comparison-scalability} and \Cref{tab:delay-diff-neigh-size}, highlight the following insights:
\begin{itemize}[leftmargin=*]
    \item \alg achieves competitive or improved performances to benchmarks, with and without delays, \textit{even though the benchmarks are tasked with easier versions of the problem that \alg solves and have higher performance bounds} (\Cref{subsec:comparison-no-delay,subsec:comparison-with-delays,subsec:scalability}).
    \item 
    \alg is observed to scale sublinearly in $|\calN|$ under delays, in contrast to  the benchmark that has almost no benefit in network growth. (\Cref{subsec:scalability}).
\end{itemize}

The common simulation setup is the same as~\Cref{sec:experiments-1}.
We next detail the benchmark algorithms (\Cref{subsec:compared-algorithms}) and comparison results without delays (\Cref{subsec:comparison-no-delay}), with delays (\Cref{subsec:comparison-with-delays}), and scalability under delays (\Cref{subsec:scalability}).
\begin{figure*}[t]
    \captionsetup{font=footnotesize}
    \centering
    \includegraphics[width=\textwidth]{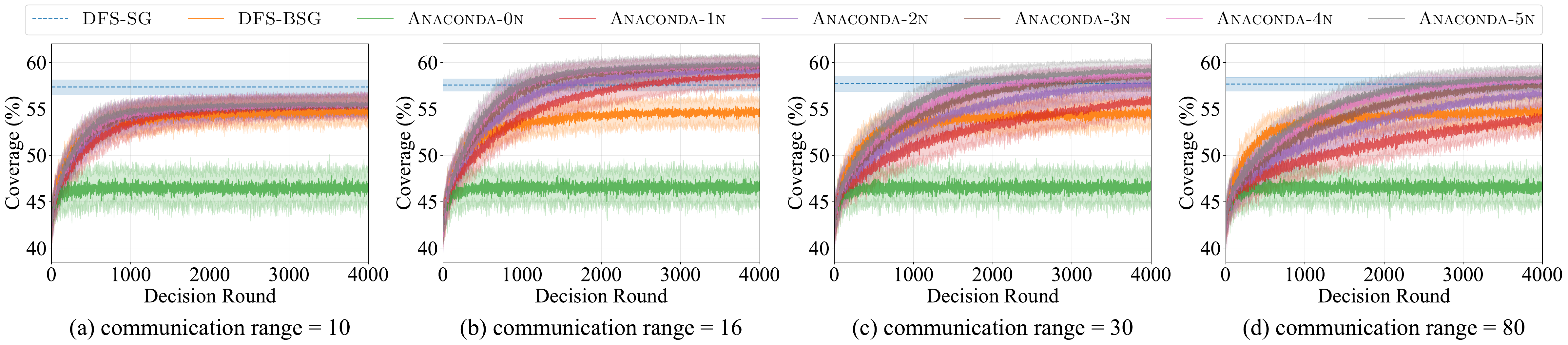}
    \vspace{-5mm}
    \caption{\textbf{Comparison of {\fontsize{8}{8}\selectfont\sc Anaconda}, {\fontsize{8}{8}\selectfont\sc DFS-SG}, and {\fontsize{8}{8}\selectfont\sc DFS-BSG} for area monitoring without computation and communication delays.} Cameras select their FOV directions using {\fontsize{8}{8}\selectfont\sc Anaconda} with maximum communication neighborhood sizes in $\{0,\dots, 5\}$, or using {\fontsize{8}{8}\selectfont\sc DFS-SG} or {\fontsize{8}{8}\selectfont\sc DFS-BSG}. From (a) to (d), the communication range $c_i$ for all cameras $i\in\calN$ increases from 10 to 16 to 20 to 80, and thus expanding each camera's coordination neighborhood $\calM_i$ growing from a small locality to the full set $\calN\setminus\{i\}$. {\fontsize{8}{8}\selectfont\sc DFS-SG} is executed for a single decision round, whereas the other algorithms are run for 4000 rounds. Results are averaged over 20 Monte Carlo trials.    
    }\label{fig:comparison-no-delay}
   \vspace{-4mm}
\end{figure*}

\subsection{Compared Algorithms}\label{subsec:compared-algorithms}
We compare the following three algorithms.

\paragraph{\alg with different communication neighborhood sizes}
We test \alg varying the maximum communication neighborhood size $|\calN_{i,t}|\,\leq\alpha_i$, namely {\fontsize{10}{10}\selectfont\sc Anaconda-$\alpha_i$n}, where $\alpha_i\in\{0,\dots, 5\}$. 

\paragraph{\dfs that requires a (strongly) connected network and a known environment~\cite{konda2022execution}} We test \dfs in terms of its achieved objective value after one round of multi-agent decisions.
The offline algorithm is an implementation with communication specifications of \sg for submodular maximization, which uses a DFS-based method to distributively determine the next agent over the (strongly) connected communication network.
Therefore, the problem solved by \dfs is a relaxed version of \Cref{pr:online} where $f$ is known and a connected communication network is given.
\dfs enjoys the same $1/(1+\kappa_f)$ suboptimality bound as \sg~\cite{fisher1978analysis} {and has a worst-case $O(\tau_c\,|\calN|^3)$ decision time. The proof of decision time appears in Appendix~C.}
\paragraph{\scenario{DFS-BSG} that requires a (strongly) connected network} We test \scenario{DFS-BSG} in terms of its achieved objective value and convergence speed across multiple rounds of multi-agent decisions. The bandit algorithm is an implementation with communication specifications of \banalg~\cite{xu2023bandit} that, similarly to \alg, also decomposes multi-agent online decision-making into single-agent problems. But different from \alg, per \scenario{DFS-BSG}, (i) each agent $i$'s action selection reward is $f(a_{i,t}\,|\,\{a_{j,t}\}_{j\in[i-1]})$ instead of $f(a_{i,t}\,|\,\{a_{j,t}\}_{j\in\calN_{i,t}})$, $\forall t$, and (ii) computating rewards is enabled by the sequential communication over all agents in a DFS-based order.\footnote{In~\cite{xu2023bandit}, the {\fontsize{8}{8}\selectfont\sc BSG} algorithm uses {\fontsize{8}{8}\selectfont\sc Exp3$^\star$-SIX} as the single-agent algorithm that provides bounded tracking regret in dynamic environments. In this paper, although we instead consider static regret and adopt {\fontsize{8}{8}\selectfont\sc Exp3}, the sequential communication scheme and decision time of {\fontsize{8}{8}\selectfont\sc BSG} are not altered.} The problem solved by \scenario{DFS-BSG} is a relaxed version of \Cref{pr:online} where a connected communication network is given. {\scenario{DFS-BSG} enjoys the same suboptimality bound as \alg in the fully centralized case per \Cref{th:main}, and thus upon convergence, \scenario{DFS-BSG} has the same guarantee as \dfs. It requires $O\mleft[(\tau_f\,|\bar{\calV}|\,|\calN|^2\,+\,\tau_c\,|\calN|^5)\,|\bar{\calV}|\,/\,\epsilon\mright]$ time to converge in a directed network. The proofs of suboptimality guarantees and decision time appear in Appendix~D.}

When implementing the algorithms above in each MC trial, we first let each agent $i$ construct $\calM_i$ within the given communication range $c_i$. Then, while each agent actively selects neighbors $\calN_{i,t}\subseteq\calM_i$ with \alg, it will directly take $\calN_{i,t}\equiv\calM_i, \forall t$ with \dfs and \scenario{DFS-BSG}. Although \alg can accommodate arbitrary networks, we need to choose a not too small $c_i$ to ensure the resulting network is connected for \dfs and \scenario{DFS-BSG}.

\subsection{Comparison with No Delays: Coverage vs. Decision Round}\label{subsec:comparison-no-delay}
To demonstrate the benefit of information access to the suboptimality of \alg, we evaluate the above algorithms omitting computation and communication delays across four scenarios (Fig.~\ref{fig:comparison-no-delay}). Each scenario is assessed over 20 MC trials, where \dfs is executed for a single decision round and all other algorithms are run for 4000 rounds.

\myParagraph{Setup} 
    \textit{Environment:} A static $50\times 50$ map. %
    \textit{Agents:} There exist $50$ cameras. The communication ranges $c_i\in\{10, 16, 30, 80\}$ across the four scenarios. In each MC trial, the location $x_{i}$ of each camera $i\in\calN$ is uniformly sampled in $[0,50]^2$.
    \textit{Actions:} Direction $a_{i,t}$ is chosen from the 16 cardinal directions, $\forall t$, with FOV radius $r_i=8$ and AOV $\theta_i=\pi/3$.

\myParagraph{Results}  
{The simulation results are presented in Fig.~\ref{fig:comparison-no-delay}, where we observe a trade-off of \alg between centralization and decentralization, where increasing $\alpha_i$ or $c_i$ generally improves coverage after convergence but at the cost of more decision rounds. Both parameters follow the principle of diminishing marginal returns due to the submodularity of \voc: while larger $\alpha_i$ and $c_i$ values enhance the agents' information access, the gains eventually become incremental. Furthermore, larger $c_i$ values significantly increase the number of decision rounds required to converge, meaning that highly centralized configurations may underperform compared to the benchmarks in shorter time horizons. Moreover, \alg consistently achieves improved asymptotic coverage over these benchmarks as long as $\alpha_i$ or $c_i$ are not too small, suggesting that intermediate parameter values, such as $c_i=16$, offer the most effective balance between real-time convergence and long-term performance. We conjecture that \alg\!\!’s improved performance arises from richer information mixing enabled by the time-varying communication neighborhoods, as opposed to the benchmarks’ sequential information passing.
}

\begin{figure*}[t!]
    \captionsetup{font=footnotesize}
	\begin{center}
	\hspace{-9.6cm}
      \begin{minipage}{\columnwidth}
            \begin{tabular}{p{\introFigColWidth}p{\introFigColWidth}p{\introFigColWidth}}
            \begin{minipage}{\introFigColWidth}%
                  \centering%
                  \includegraphics[width=\columnwidth]{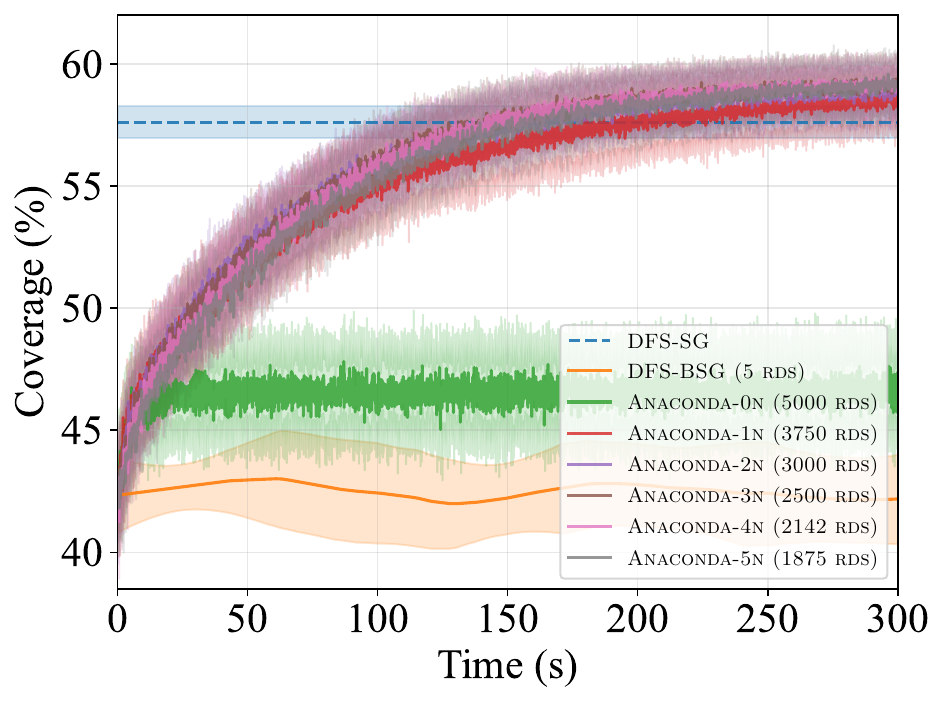} \\
                  \caption*{(a) {$\tau_f = .01$s, $\tau_c = .03$s.}
                  }
            \end{minipage}
            &
            \begin{minipage}{\introFigColWidth}%
                  \centering%
                  \includegraphics[width=\columnwidth]{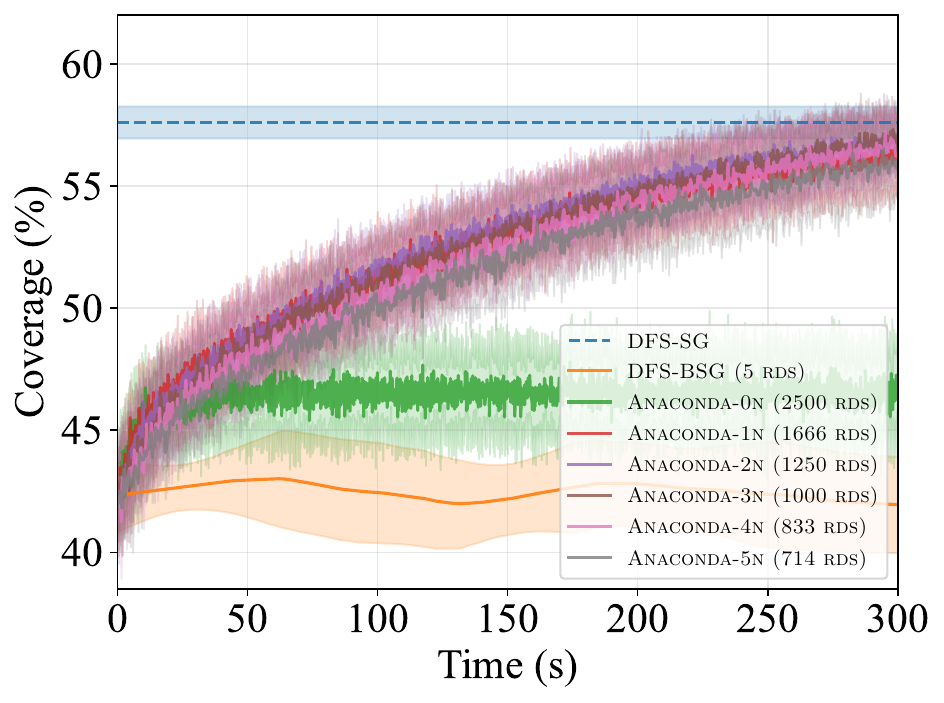} \\ 
                  \caption*{(b) {$\tau_f = .03$s, $\tau_c = .03$s.}
                  }
            \end{minipage}
            &
            \begin{minipage}{\introFigColWidth}%
                  \centering%
                  \includegraphics[width=\columnwidth]{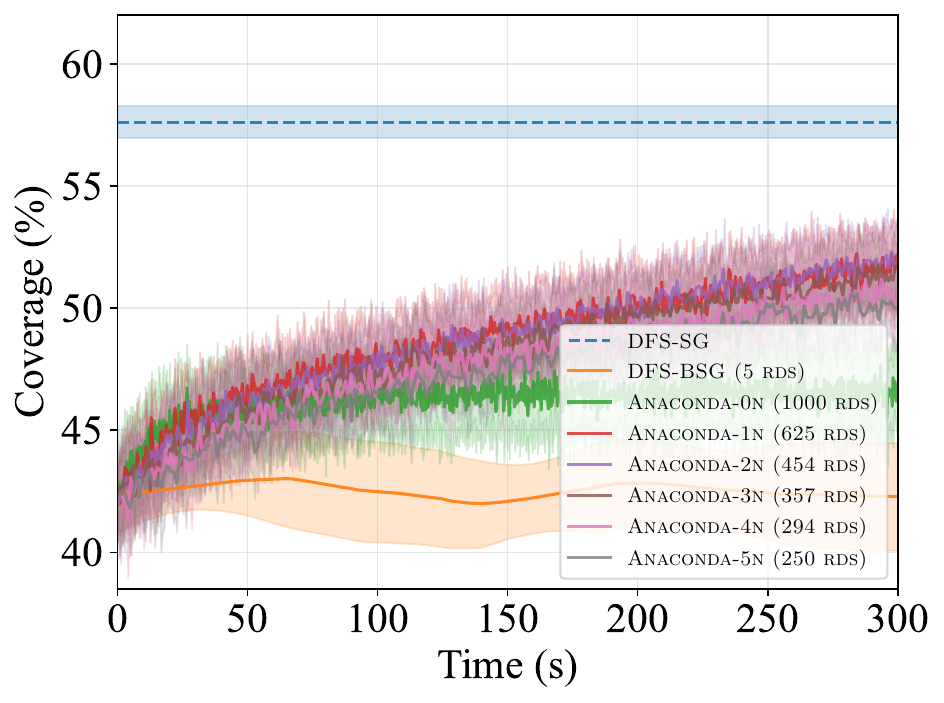} \\
                  \caption*{
                  (c) {$\tau_f = .09$s, $\tau_c = .03$s.}
                  }
            \end{minipage}
            \end{tabular}
	\end{minipage} 
	\caption{\textbf{Comparison of {\fontsize{8}{8}\selectfont\sc Anaconda} vs. {\fontsize{8}{8}\selectfont\sc DFS-SG} vs. {\fontsize{8}{8}\selectfont\sc DFS-BSG} for real-time coverage performance under computation and communication delays.} Cameras select their FOV directions using {\fontsize{8}{8}\selectfont\sc Anaconda} with maximum communication neighborhood sizes in $\{0,\dots, 5\}$, or using {\fontsize{8}{8}\selectfont\sc DFS-SG} or {\fontsize{8}{8}\selectfont\sc DFS-BSG}. From (a) to (b) to (c), the time $\tau_f$ for one function evaluation increases relative to the delay $\tau_c$ for transmitting one action through a communication link, with the ratio $\tau_f/\tau_c$ taking values $1/3$, $1$, and $3$. {\fontsize{8}{8}\selectfont\sc DFS-SG} is executed for a single decision round, whereas the other algorithms are run for a fixed duration of $300$ seconds. Under different delay configurations, the algorithms complete different numbers of decision rounds within this time window. Results are averaged over 20 Monte Carlo trials. 
	}\label{fig:comparison-with-delay}
	\vspace{-6mm}
	\end{center}
\end{figure*}

\subsection{Comparison with Delays: Coverage vs. Actual Time}\label{subsec:comparison-with-delays}

We compare \alg with \dfs and \scenario{DFS-BSG} in 20 MC trials under three delay configurations $(\tau_f,\tau_c)\in\{(.01\text{s},.03\text{s}),(.03\text{s},.03\text{s}),(.09\text{s},.03\text{s})\}$, capturing different computation and communication capabilities. Each trial is carried out over a fixed $300$s time horizon.

\myParagraph{Setup} The setup is identical to that in Fig.~\ref{fig:comparison-no-delay}(b), with the communication range $c_i=16$ for all cameras $i\in\calN$.

\myParagraph{Results} 
The simulation results are presented in Fig.~\ref{fig:comparison-with-delay} and \Cref{tab:delay-diff-neigh-size} where we observe the following for \alg:
\alg demonstrates a better performance than \scenario{DFS-BSG} and can outperform \scenario{DFS-SG} after convergence. The reason is that \alg decides actions much faster than \scenario{DFS-BSG} in large networks, which means much more decision rounds in a fixed time horizon. 

Increasing the neighborhood size $\alpha_i$ presents a trade-off for \alg: although larger neighborhoods theoretically offer higher asymptotic coverage, they also increase computation time per round, resulting in fewer decision cycles within a fixed time horizon. For example, in Fig.~\ref{fig:comparison-with-delay}(a)–(c), the best-performing communication neighborhood sizes within the 300s window are $\alpha_i=5, 3,$ and $1$, respectively, rather than always $5$ (\Cref{tab:delay-diff-neigh-size}). This ``no free neighbor'' dynamic, characterized by a cubic growth in convergence time relative to $\alpha_i$, means that smaller neighborhood sizes can yield better performance in a fixed horizon.}

\subsection{Scalability}\label{subsec:scalability}
We finally compare the real-time coverage performance of {\fontsize{10}{10}\selectfont\sc Anaconda-5n}  and \scenario{DFS-BSG} over 20 MC trials as the network size scales across five scenarios with varying numbers of cameras and map sizes. Each trial is executed over a fixed time horizon of $300$s with $(\tau_f,\tau_c) = (0.01\text{s}, 0.01\text{s})$.

\begin{table}[tbp]
\captionsetup{font=footnotesize}
\centering
\caption{Comparison of average coverage performance (\%) within $300$s across three scenarios with different delay configurations (Fig.~\ref{fig:comparison-with-delay}). The highest value in each scenario is highlighted in \textbf{bold}. 
}
\scriptsize
\setlength{\tabcolsep}{4pt}
\resizebox{\columnwidth}{!}{%
\begin{tabular}{@{}c|c|c|c@{}}
\toprule
\multirow{2}{*}{\textbf{Algorithm}} & \multicolumn{3}{c@{}}{\textbf{Coverage (\%)}} \\
\cmidrule(l){2-4}
 & $\tau_f{=}.01s$, $\tau_c{=}.03s$ & $\tau_f{=}.03s$, $\tau_c{=}.03s$ & $\tau_f{=}.09s$, $\tau_c{=}.03s$ \\
\midrule
{\scriptsize \sc DFS-BSG} & 42.19 $\pm$ 1.83 & 41.96 $\pm$ 1.98 & 42.30 $\pm$ 2.20 \\

{\scriptsize \sc Anaconda-0n} & 47.33 $\pm$ 1.84 & 46.52 $\pm$ 1.92 & 46.78 $\pm$ 1.77 \\

{\scriptsize \sc Anaconda-1n} & 58.27 $\pm$ 1.22 & 56.13 $\pm$ 1.77 & \textbf{52.17 $\pm$ 1.68} \\

{\scriptsize \sc Anaconda-2n} & 58.75 $\pm$ 1.18 & 56.75 $\pm$ 1.28 & 51.68 $\pm$ 1.90 \\

{\scriptsize \sc Anaconda-3n} & 59.05 $\pm$ 0.89 & \textbf{57.10 $\pm$ 1.04} & 51.73 $\pm$ 0.91 \\

{\scriptsize \sc Anaconda-4n} & 59.08 $\pm$ 0.82 & 56.20 $\pm$ 1.57 & 50.12 $\pm$ 0.93 \\

{\scriptsize \sc Anaconda-5n} & \textbf{59.12 $\pm$ 1.01} & 55.72 $\pm$ 1.32 & 49.99 $\pm$ 1.73 \\
\bottomrule
\end{tabular}%
}
\label{tab:delay-diff-neigh-size}
\end{table}

\myParagraph{Setup} The five scenarios contain $\{10, 20, 30, 40, 50\}$ cameras deployed in maps of sizes $\{23^2, 32^2, 39^2, 45^2, 50^2\}$, respectively. These configurations maintain an approximately constant camera density, with each camera covering roughly $50$ square units on average. Across all scenarios, the communication range is fixed at $c_i=16$ for all cameras $i\in\calN$.

\begin{figure}[t]
    \captionsetup{font=footnotesize}
    \centering
    \includegraphics[width=\columnwidth]{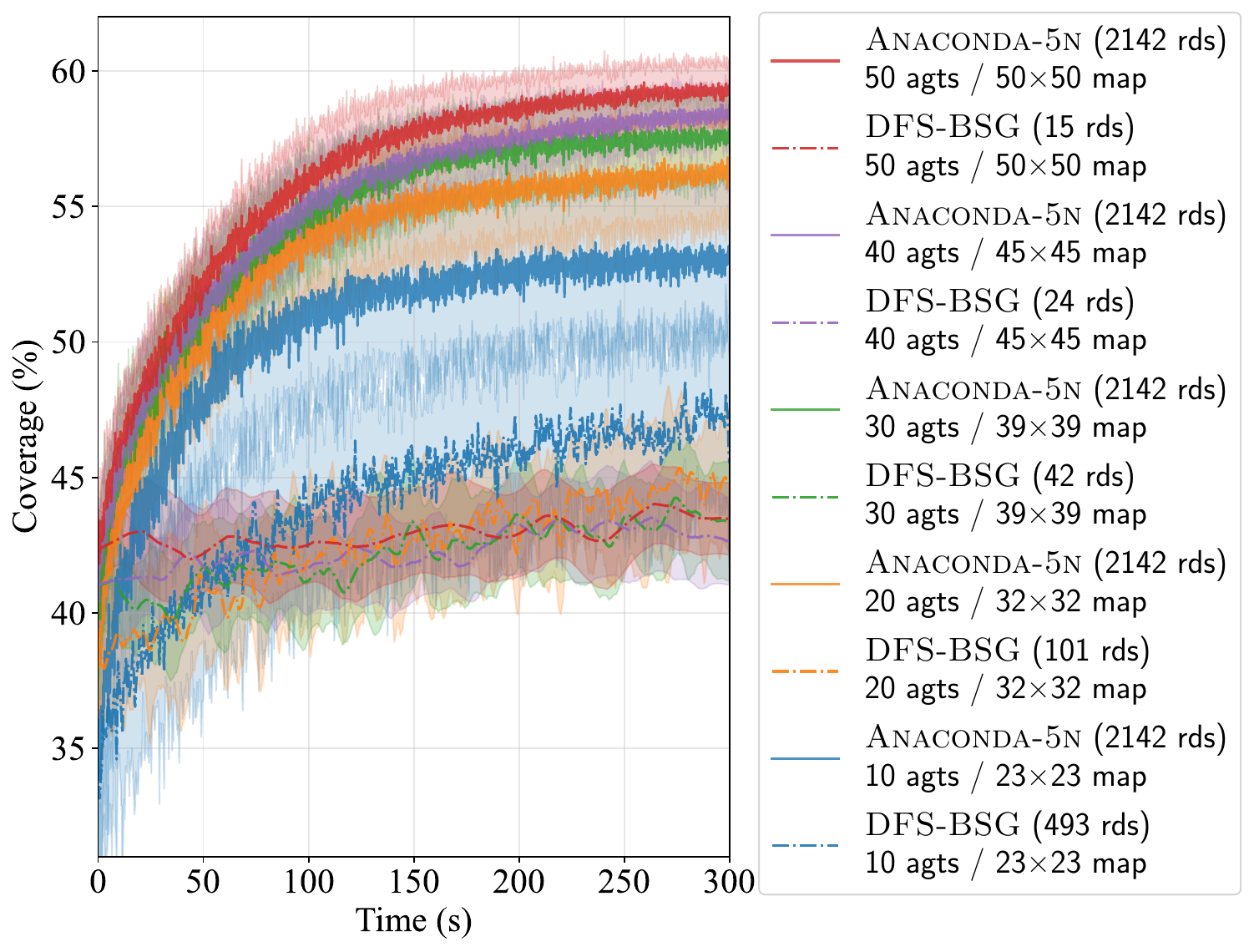}
    \vspace{-3mm}
    \caption{\textbf{Comparison of {\fontsize{8}{8}\selectfont\sc Anaconda-5n} vs. {\fontsize{8}{8}\selectfont\sc DFS-BSG} in real-time coverage performance with scaling network.} Cameras select their FOV directions using {\fontsize{8}{8}\selectfont\sc Anaconda-5n} or {\fontsize{8}{8}\selectfont\sc DFS-BSG}, across five scenarios with different network and map sizes. To keep the camera density constant, the map size scales linearly from $23\times 23$ to $50\times 50$ as the network size ranges from $10$ to $50$. Results are averaged over 20 Monte Carlo trials, each with a fixed $300$s time window under delays $(\tau_f, \tau_c) = (.01\text{s}, .01\text{s})$. 
    }\label{fig:comparison-scalability}
\end{figure}

\myParagraph{Results} 
{{\alg scales despite imperfect communication; \scenario{DFS-BSG} does not}}.
    Per Fig.~\ref{fig:comparison-scalability}, when the network size increases from 10 to 50 agents, \alg consistently executes 2142 rounds, owing to its fixed per-round computation and communication load (\Cref{prop:decision-time}), and the convergence time grows from 100s to 200s, which is a sublinear growth in $|\calN|$. In contrast, the number of rounds completed by \scenario{DFS-BSG} declines from 493 to just 15, because each round requires sequential communication across the entire network.

\section{Conclusion} \label{sec:con}
We presented \alg, a scalable framework for distributed submodular coordination in multi-agent systems operating in unknown environments under realistic communication constraints. \alg achieves scalability while maintaining near-optimal action coordination by intelligently limiting communication, regardless of global connectivity. %
We introduced a novel metric, {\fontsize{9}{9}\selectfont\sf VoC}, that quantifies the benefit of information access for coordination. By optimizing  {\fontsize{9}{9}\selectfont\sf VoC}, the framework enables intelligent information limitation through communication neighborhood design. 
The suboptimality guarantees showed that \alg's coordination performance improves with the sum of all agents' {\fontsize{9}{9}\selectfont\sf VoC}, and is anytime non-trivial even prior to convergence.
Extensive simulations in multi-camera area monitoring against state-of-the-art benchmarks supported the theoretical analysis and illustrated \alg's benefit in coordination performance, neighborhood design, and scalability, together with a fundamental trade-off between coordination optimality and convergence speed under realistic communication latency.

\myParagraph{Future work}
While \alg demonstrates improvement in decision time compared to benchmarks for unknown environments, for mobile-robot applications, we will work to accelerate it to $O(|\calN|)$ decision time, leveraging the tools in~\cite{rakhlin2013online}.
Moreover, while the coordination neighborhood $\calM_i$ is currently considered static, it can be dynamic for mobile robots. To this end, we will also handle dynamic coordination neighborhoods, \ie $\calM_{i,t}$, with guaranteed performances.

\appendices

\section{Suboptimality Guarantees of \alg}\label{app:main}
We will first prove \Cref{lem:SubMI,lem:AReg-bound,lem:NReg-bound}, then \Cref{prop:action} and \Cref{th:main,th:posteriori,th:asymptotic}.

\myParagraph{Proof of \Cref{lem:SubMI}%
}%
Consider $\smi{a}{\calJ}= f(a) - f(a\,|\,\{a_{j}\}_{j\myin\calJ})$,  where $\calJ\subseteq\calM_i\subseteq\calN\setminus\{i\}$, $a\in\calV_i$ is fixed, and $f\colon2^{\calV_\calN}\mapsto \mathbb{R}$ is non-decreasing and 2nd-order submodular. Also, with a slight abuse of notation, denote $f(a\,|\,\{a_{j}\}_{j\myin\calJ})$ by $f(a\,|\,\calJ)$. 

To prove the monotonicity of $\smi{a}{\cdot}$, consider $\calA_1, \calA_2\subseteq\calM_i$ that are disjoint.  Then,
$\smi{a}{\calA_1\cup\calA_2} - \smi{a}{\calA_1} = -f(a\,|\,\calA_1\cup\calA_2) + f(a\,|\,\calA_1)\geq 0$ since $f$ is submodular.  Thus, $\smi{a}{\cdot}$ is non-decreasing. 

To prove the submodularity of $\smi{a}{\cdot}$, consider $\calA, \calB_1, \calB_2\subseteq\calV$, where $\calB_1$ and $\calB_2$ are disjoint, then:
{%
\begin{align}
   & \hspace{-1.5mm}\smi{a}{\calA\,|\,\calB_1} - \smi{a}{\calA\,|\,\calB_1\cup\calB_2}\nonumber\\
    & \hspace{-2mm}=\smi{a}{\calA\cup\calB_1} - \smi{a}{\calB_1}\nonumber\\
    & \hspace{-2mm}\hspace{1.5cm}- \smi{a}{\calA\cup\calB_1\cup\calB_2} + \smi{a}{\calB_1\cup\calB_2}\nonumber\\
    & \hspace{-2mm}=-f(a\,|\,\calA\cup\calB_1) + f(a\,|\,\calB_1)\nonumber\\
    & \hspace{-2mm}\hspace{1.5cm} + f(a\,|\,\calA\cup\calB_1\cup\calB_2) - f(a\,|\,\calB_1\cup\calB_2)\geq 0,
\end{align}}where the inequality holds since $f$ is 2nd-order submodular (\Cref{def:conditioning}). Therefore, $\smi{a}{\cdot}$ is submodular. \qed

\myParagraph{Proof of~\Cref{lem:AReg-bound}%
}
    According to~\cite[Theorem 3.1]{auer2002nonstochastic}, we have
    $\mathbb{E}\mleft[\operatorname{A-Reg}_T\mleft(\{a_{i,t}\}_{t\in [T]}\mright)\mright] \leq \sqrt{2\,T\,|\calV_i|\,\log{|\calV_i|}}$. 
    \qed

\myParagraph{Proof of~\Cref{lem:NReg-bound}%
}
Given~\Cref{def:NReg}, we have
{\allowdisplaybreaks
\begin{align}
    &\mathbb{E}\mleft[\operatorname{N-Reg}_{\{a_{i,t}\}_{t\in [T]}}^{\rho(\kappa_{I,i}, \alpha_i)}\mleft(\{\calN_{i,t}\}_{t\myin [T]}\mright)\mright] \nonumber    \\ %
    &= \rho(\kappa_{I,i}, \alpha_i) \max_{{\mathcal{N}_{i,t} \subseteq \mathcal{M}_i, |\mathcal{N}_{i,t}| \leq \alpha_i}}\sum_{t=1}^{T} \smi{a_{i,t}}{\calN_{i,t}} \nonumber \\
    & \quad- \mathbb{E}\mleft[\sum_{t=1}^{T}\smi{a_{i,t}}{\calN_{i,t}}\mright]\nonumber \\\label{aux2:2}
    &\leq \rho(\kappa_{I,i}, \alpha_i) \sum_{t=1}^{T} \mathbb{E} \mleft[ - \alpha_i\, r_{j_{k,t}, t}+\sum_{k=1}^{\alpha_i}r_{j_{k,t}^\opt, t}\mright]\\\label{aux2:3}
    &= \rho(\kappa_{I,i}, \alpha_i) \sum_{k=1}^{\alpha_i}\mathbb{E} \mleft[\sum_{t=1}^{T} \mleft(r_{j_{k,t}^\opt, t} - r_{j_{k,t}, t}^{\top}\, q_{k,t}\mright)\mright]\\\label{aux2:5}
    &\leq\Tilde{O}\mleft(\alpha_i\sqrt{|\calM_i|T}\mright),
\end{align}}where %
\cref{aux2:2} follows from \cite[Theorem 3]{matsuoka2021tracking}, \cref{aux2:3} follows from the linearity of expectation, and \cref{aux2:5} follows by applying \cite[Theorem 3.1]{auer2002nonstochastic}. \qed %

\myParagraph{Proofs of \Cref{prop:action} and \Cref{th:main}}%
We have:
{\allowdisplaybreaks\begin{align}
    &\sum_{t=1}^{T}f(\calA^\opt) \nonumber\\
    &=\sum_{t=1}^{T} f(\calA^\opt\cup\calA_t) - \sum_{t=1}^{T} \sum_{i\in\calN} f(a_{i,t}\,|\,\calA^\opt\cup\{a_{j,t}\}_{j\in[i-1]}) \label{aux22:1}\\
    &\leq \sum_{t=1}^{T} f(\calA_t) + \sum_{t=1}^{T} \sum_{i\in\calN} f(a_{i}^\opt\,|\,\calA_t) \nonumber\\
    &\quad - (1-\kappa_{f}) \sum_{t=1}^{T} \sum_{i\in\calN} f(a_{i,t}\,|\,\{a_{j,t}\}_{j\in\calN_{i,t}}) \label{aux22:2}\\
    &\leq \sum_{t=1}^{T} f(\calA_t) + \kappa_{f} \sum_{t=1}^{T} \sum_{i\in\calN} f(a_{i,t}\,|\,\{a_{j,t}\}_{j\in\calN_{i,t}}) \label{aux22:3}\\\nonumber
    &\quad + \sum_{i\in\calN} \sum_{t=1}^{T} \mleft[f(a_{i}^\opt\,|\,\{a_{j,t}\}_{j\in\calN_{i,t}}) - f(a_{i,t}\,|\,\{a_{j,t}\}_{j\in\calN_{i,t}})\mright] \\  
    &\leq \sum_{t=1}^{T} f(\calA_t) + \kappa_{f} \sum_{t=1}^{T} \sum_{i\in\calN} f(a_{i,t}\,|\,\{a_{j,t}\}_{j\in\calN_{i,t}}) \nonumber\\
    &\quad + \sum_{i\in\calN} \AReg\mleft(\{a_{i,t}\}_{t\in [T]}\mright) \label{aux22:4}\\
    &= \sum_{t=1}^{T} f(\calA_t) - \kappa_{f}\sum_{i\in\calN} \sum_{t=1}^{T} \smi{a_{i,t}}{\calN_{i,t}} \nonumber\\
    & \quad + \kappa_{f} \sum_{i\in\calN} \sum_{t=1}^{T} f(a_{i,t}) + \sum_{i\in\calN} \AReg\mleft(\{a_{i,t}\}_{t\in [T]}\mright) \label{aux22:5}\\
    &\leq \sum_{t=1}^{T} f(\calA_t) + \frac{\kappa_{f}}{1-\kappa_{f}} \sum_{t=1}^{T} \sum_{i\in\calN} f(a_{i,t}\,|\,\{a_{j,t}\}_{j\in[i-1]}) \label{aux22:55} \\
    & \quad- \kappa_{f}\sum_{i\in\calN} \sum_{t=1}^{T} \smi{a_{i,t}}{\calN_{i,t}}  + \sum_{i\in\calN} \AReg\mleft(\{a_{i,t}\}_{t\in [T]}\mright) \nonumber\\
    &\leq \frac{1}{1-\kappa_{f}}\sum_{t=1}^{T} f(\calA_t) + \kappa_{f}\sum_{i\in\calN} \operatorname{N-Reg}_{\{a_{i,t}\}_{t\in [T]}}^{\rho(\kappa_{I}, \bar{\alpha})}\mleft(\{\calN_{i,t}\}_{t\in [T]}\mright) \nonumber \\
    & \quad- \kappa_f \,\rho(\kappa_{I}, \bar{\alpha}) \sum_{i\in\calN} \sum_{t=1}^{T} \smi{a_{i,t}}{\calN_{i}^{\star}(\{a_{i,t}\}_{t\in [T]}; \alpha_i, \calM_i)} \nonumber \\
    &\quad + \sum_{i\in\calN} \AReg\mleft(\{a_{i,t}\}_{t\in [T]}\mright), \label{aux22:6}
\end{align}}where \cref{aux22:1} holds by telescoping the sum, \cref{aux22:2} holds since $f$ is submodular and since $1-\kappa_{f} \leq \frac{f(a_{i,t}\,|\,\calA^\opt\,\cup\,\{a_{j,t}\}_{j\in\calN\setminus\{i\}})}{f(a_{i,t})} \leq \frac{f(a_{i,t}\,|\,\calA^\opt\,\cup\,\{a_{j,t}\}_{j\in[i-1]})}{f(a_{i,t}\,|\,\{a_{j,t}\}_{j\in\calN_{i}})}$ per \Cref{def:curvature}, \cref{aux22:3} holds from submodularity, \cref{aux22:4} holds from \Cref{def:action-regret}, \cref{aux22:5} holds from \Cref{def:MI}, \cref{aux22:55} holds since $1-\kappa_{f} \leq \frac{f(a_{i,t}\,|\,\calA^\opt\,\cup\,\{a_{j,t}\}_{j\in\calN\setminus\{i\}})}{f(a_{i,t})} \leq \frac{f(a_{i,t}\,|\,\{a_{j,t}\}_{j\in[i-1]})}{f(a_{i,t})}$ per \Cref{def:curvature}, and \cref{aux22:6} holds from \Cref{def:NReg}. After taking expectations on both sides of \cref{aux22:55}, with \Cref{lem:AReg-bound}, we prove \Cref{prop:action}.

Then, taking expectations on both sides of \cref{aux22:6}, with \Cref{lem:AReg-bound,lem:NReg-bound}, we have
{\allowdisplaybreaks\begin{align}
    &T f(\calA^\opt) \leq \frac{1}{1-\kappa_f} \mathbb{E}\mleft[\sum_{t=1}^{T}f(\calA_t)\mright] \nonumber\\
    &\quad - \kappa_f \,\rho(\kappa_{I}, \bar{\alpha})\nonumber \\
    &\qquad\times\sum_{i\in\calN} \mathbb{E}\mleft[\sum_{t=1}^{T} \smi{a_{i,t}}{\calN_{i}^{\star}(\{a_{i,t}\}_{t\in [T]}; \alpha_i, \calM_i)}\mright] \nonumber \\
    &\quad + \Tilde{O}\mleft(|\calN|\bar{\alpha}\sqrt{{|\bar{\calM}|}{T}}\mright) + \Tilde{O}\mleft(|\calN|\sqrt{{|\bar{\calV}|}{T}}\mright). \label{aux22:7} 
\end{align}}%

Reorganizing \cref{aux22:7}, we prove \cref{eq:thm-3}:
\begin{align}
    &\mathbb{E}\mleft[\sum_{t=1}^{T}f(\calA_t)\mright] \geq (1-\curv_f)\, T f(\solopt) \nonumber\\
    &\quad+ \kappa_f\, (1-\kappa_f)\, \rho(\kappa_{I}, \bar{\alpha}) \nonumber \\
    &\qquad\times \sum_{i\in\calN} \mathbb{E}\mleft[\sum_{t=1}^{T}\smi{a_{i,t}}{\calN_{i}^{\star}(\{a_{i,t}\}_{t\in [T]}; \alpha_i, \calM_i)}\mright] \nonumber\\
    &\quad- \Tilde{O}\mleft(|\calN|\sqrt{{\mleft(\bar{\alpha}^2\,|\bar{\calM}|\,+\,|\bar{\calV}|\mright)}{T}}\mright).
\end{align}

{%
To prove \cref{eq:thm-1}, where $\calN_{i,t}\equiv\calM_i=\calN\setminus\{i\}$, we take expectations on both sides of \cref{aux22:4}: 
\begin{align}
    &\mathbb{E}\mleft[\sum_{t=1}^{T}f(\calA_t)\mright] \geq T f(\solopt)\nonumber \\
    &\quad- \kappa_f \sum_{i\in\calN} \mathbb{E}\mleft[\sum_{t=1}^{T}f(a_{i,t}\,|\,\calA_{t}\setminus\{a_{i,t}\})\mright]- \Tilde{O}\mleft(|\calN|\sqrt{|\bar{\calV}|T}\mright)\label{aux2222:1}\\
    &\geq T f(\solopt) - \kappa_f\, \mathbb{E}\mleft[\sum_{t=1}^{T}f(\calA_t)\mright] - \Tilde{O}\mleft(|\calN|\sqrt{|\bar{\calV}|T}\mright),\label{aux2222:2}
\end{align}}where \cref{aux2222:1} holds from \cref{aux22:4}, and \cref{aux2222:2} holds from~\cite[Eq.~(15)]{tzoumas2017resilient}. Thereby, 

\begin{equation}
    \hspace{-1.5mm}\mathbb{E}\mleft[\sum_{t=1}^{T}f(\calA_t)\mright] \geq \frac{1}{1+\kappa_f} T f(\solopt) - \Tilde{O}\mleft(|\calN|\sqrt{|\bar{\calV}|T}\mright).
\end{equation}

To prove \cref{eq:thm-2}, where $\calN_{i,t}\equiv\emptyset$, we, again, take expectations on both sides of \cref{aux22:4}:
\begin{align}
    &\hspace{-1mm}\mathbb{E}\mleft[\sum_{t=1}^{T}f(\calA_t)\mright]\nonumber\\
    &\hspace{-1mm}\geq T f(\solopt) - \kappa_f \sum_{i\in\calN} \mathbb{E}\mleft[\sum_{t=1}^{T}f(a_{i,t})\mright] - \Tilde{O}\mleft(|\calN|\sqrt{|\bar{\calV}|T}\mright)\nonumber\\
    &\hspace{-1mm}\geq T f(\solopt) - \frac{\kappa_f}{1-\kappa_f}\, \mathbb{E}\mleft[\sum_{t=1}^{T}f(\calA_t)\mright] - \Tilde{O}\mleft(|\calN|\sqrt{|\bar{\calV}|T}\mright).
\end{align}
Therefore, 

\begin{equation}
    \hspace{-1.5mm}\mathbb{E}\mleft[\sum_{t=1}^{T}f(\calA_t)\mright] \geq (1-\curv_f) T f(\solopt) - \Tilde{O}\mleft(|\calN|\sqrt{|\bar{\calV}|T}\mright).
\end{equation}
\qed

\myParagraph{Proof of \Cref{th:posteriori}}%

Consider neighborhoods $\{\calN_{i,t}^{\mathsf{DAG}}\}_{i\in\calN}$ are induced by $\calG_t^{\mathsf{DAG}}$, then $\calN_{i,t}^{\mathsf{DAG}}\subseteq\calN_{i,t}, \forall i\in\calN$. Dividing both sides of \cref{aux22:4} by $\sum_{t=1}^{T} f(\calA_t)$, we have
\begin{align}
    \frac{T f(\calA^\opt)}{\sum_{t=1}^{T}f(\calA_t)} &\leq 1 + \kappa_{f} \underbrace{\frac{\sum_{t=1}^{T} \sum_{i\in\calN} f(a_{i,t}\,|\,\{a_{j,t}\}_{j\in\calN_{i,t}})}{\sum_{t=1}^{T}f(\calA_t)}}_{\beta\text{: computable after \Cref{alg:main} terminates}} \nonumber\\
    &\quad + \frac{\sum_{i\in\calN} \AReg\mleft(\{a_{i,t}\}_{t\in [T]}\mright)}{\sum_{t=1}^{T}f(\calA_t)}. \label{aux222:1}
\end{align}Taking expectations on both sides of \cref{aux222:1},
\begin{align}%
    \hspace{-.1cm}\mathbb{E}\mleft[\frac{T f(\calA^\opt)}{\sum_{t=1}^{T}f(\calA_t)}\mright] &\leq 1 + \kappa_{f}\, \beta + \sum_{i\in\calN} \mathbb{E}\mleft[\frac{\AReg\mleft(\{a_{i,t}\}_{t\in [T]}\mright)}{\sum_{t=1}^{T}f(\calA_t)}\mright] \nonumber\\
    &\hspace{-1.2cm}\leq 1 + \kappa_{f}\, \beta + \sum_{i\in\calN} \frac{\mathbb{E}\mleft[\AReg\mleft(\{a_{i,t}\}_{t\in [T]}\mright)\mright]}{T\, \min_{t\in[T]} f(\calA_t)} \label{aux222:3}\\
    &\hspace{-1.2cm}\leq 1 + \kappa_{f}\, \beta + \Tilde{O}\mleft(|\calN|\sqrt{{|\bar{\calV}|}/{T}}\mright),\label{aux222:4}
\end{align}where \cref{aux222:3} holds since $f$ is normalized, \ie $f(\calA_t)>0$ when $\calA_t\neq\emptyset$, and \cref{aux222:4} holds from \Cref{lem:AReg-bound}. Therefore, 
\begin{align}
    \mathbb{E}\mleft[\frac{\sum_{t=1}^{T}f(\calA_t)}{T f(\calA^\opt)}\mright] &\geq {1} \Bigg/ {\mathbb{E}\mleft[\frac{T f(\calA^\opt)}{\sum_{t=1}^{T}f(\calA_t)}\mright]} \label{aux222:5}\\
    &\geq \frac{1}{1 + \kappa_{f}\, \beta + \Tilde{O}\mleft(|\calN|\sqrt{{|\bar{\calV}|}/{T}}\mright)},\label{aux222:6}
\end{align}where \cref{aux222:5} holds from Jensen's Inequality, and \cref{aux222:6} holds from \cref{aux222:4}. Moreover, \cref{eq:beta} is proved by
\begin{align}
    0 \leq \beta &\leq \frac{\sum_{t=1}^{T} \sum_{i\in\calN} f(a_{i,t}\,|\,\{a_{j,t}\}_{j\in\calN_{i,t}^{\mathsf{DAG}}})}{\sum_{t=1}^{T}f(\calA_t)}\nonumber\\
    &\leq \frac{\sum_{t=1}^{T} \alpha^\star(\calG_{t}^{\mathsf{DAG}})\, f(\calA_t)}{\sum_{t=1}^{T}f(\calA_t)} \leq \bar\alpha^\star,
\end{align}where the first inequality holds since $f$ is submodular and $\calN_{i,t}^{\mathsf{DAG}}\subseteq\calN_{i,t}, \forall i\in\calN$, and the second inequality holds from \cite[Theorem 1]{grimsman2019impact}. Hence, \Cref{th:posteriori} holds. 

In particular, if at any time $t$ there exists an agent ordering such that $[i-1] \subseteq \calN_{i,t}, \forall i$, then
\begin{align}
    0 \leq \beta &= \frac{\sum_{t=1}^{T} \sum_{i\in\calN} f(a_{i,t}\,|\,\{a_{j,t}\}_{j\in\calN_{i,t}})}{\sum_{t=1}^{T}f(\calA_t)} \nonumber\\
    & \leq \frac{\sum_{t=1}^{T} \sum_{i\in\calN} f(a_{i,t}\,|\,\{a_{j,t}\}_{j\in [i-1]})}{\sum_{t=1}^{T}f(\calA_t)} = 1
\end{align}holds due to submodularity. \qed

\myParagraph{Proof of \Cref{th:asymptotic}}
Finally, combining \cref{eq:thm-2,eq:asymp-posteriori}, \Cref{th:asymptotic} holds. \qed

\section{Decision Time of \alg}\label{app:speed}
We will first present and prove the following propositions and then prove \Cref{th:speed}.

\begin{proposition}[Convergence Rate]\label{prop:convergence-rate}
    \alg's convergence error takes $T$ iterations to be within $\epsilon$ where 
\begin{itemize}[leftmargin=*]
    \item If \neighborsel is not involved, \ie $\forall i$, $\alpha_i=0$ or $\alpha_i\geq|\calM_i|$, 
    \begin{equation}
        T\geq |\bar{\calV}||\calN|^2\,/\,\epsilon,
    \end{equation}
    \item If \neighborsel is involved, \ie $\exists i\in\calN$, $0<\alpha_i<|\calM_i|$, 
        \begin{equation}T\geq \mleft(\bar{\alpha}^2\,|\bar{\calM}|\,+\,\bar{|\calV|}\mright)\,|\calN|^2\,/\,\epsilon.
        \end{equation}
\end{itemize}
\end{proposition}
\paragraph*{Proof} \Cref{prop:convergence-rate} holds from \Cref{lem:AReg-bound,lem:NReg-bound}. \qed %

\begin{proposition}[Computational Complexity]\label{prop:computation}
At each $t\in [T]$, \alg requires each agent $i$ to execute $2\alpha_i+3$ evaluations of $f$ and $O(|\calV_i|+\alpha_i|\calM_i|)$ additions/multiplications.
\end{proposition}

\paragraph*{Proof} At each $t\in[T]$, \actionsel requires $2$ function evaluations (\Cref{alg:action}'s line 7), along with $O(|\calV_i|)$ additions and multiplications  (\Cref{alg:action}'s lines 4 and 8). Also, \neighborsel requires $2\alpha_i+1$ function evaluations  (\Cref{alg:neighbor}'s line 9), along with $O(\alpha_i|\calM_i|)$ additions and multiplications (\Cref{alg:neighbor}'s lines 6 and 9-11). \qed

\begin{proposition}[Communication Complexity]\label{prop:communication}
At each $t\myin [T]$, \alg requires~one  communication round where each agent $i$ only transmits its own action to its out-neighbors.
\end{proposition}
\paragraph*{Proof} At each $t\in[T]$, \alg requires one (multi-channel) communication round where each agent $i$ shares $a_{i,t}$ with and simultaneously receives $\{a_{j, t}\}_{j\in\calN_{i,t}}$ from $\calN_{i,t}$ (\Cref{alg:main}'s line 5).

\begin{proposition}[Per-Round Decision Time]\label{prop:decision-time}
One round of \alg takes $\tau_f\,(2 \alpha_i+3) \,+\,\tau_c$ time to complete.
\end{proposition}
\paragraph*{Proof} \Cref{prop:decision-time} holds because of \Cref{prop:computation,prop:communication}, ignoring the time for additions and multiplications.    \qed

Finally, we prove \Cref{th:speed}:

\myParagraph{Proof of \Cref{th:speed}}
\Cref{th:speed} holds because of \Cref{prop:convergence-rate,prop:decision-time}. \qed

\section{Worst-Case Decision Time of Sequential Communication in Directed Networks}

We prove that \dfs has a $O\mleft(\tau_c\,|\calN|^3\mright)$ worst-case communication time on a strongly connected directed graph. The proof extends \cite[Section III-D]{konda2022execution} by taking also the size of each inter-agent communication message into consideration since the larger the size the more time it will take for the message to be transmitted. We use the notation:
\begin{itemize}[leftmargin=3.5mm]
    \item $\calG_{dir}=\{\calN, \calE_{dir}\}$ is a strongly connected directed graph;
    \item $\pi\colon\{1,\dots,|\calN|\}\mapsto\{1,\dots,|\calN|\}$ denotes the agents' order of action selection given by the DFS approach in \cite{konda2022execution};
    \item $d(i,j)$ denotes the length of the shortest path from agent $i$ to $j$ on $\calG_{dir}$.
\end{itemize}
Suppose $p=(v_1,\dots,v_l)$ is the longest path of $\calG_{dir}$, where
$l = |p|$. If $l=|\calN|$, then $p$ is a spanning walk on $\calG_{dir}$ with $\pi(i)=i, \forall i=\{1,\dots,|\calN|\}$ and 
\begin{align}
    &\max_{\calG_{dir}}T_{min}(\calG_{dir}) = \sum_{i\,=\,1}^{|\calN|-1} i\tau_c \times d(i,i+1) \nonumber\\
    &=\sum_{i\,=\,1}^{|\calN|-1} i\tau_c\times 1 =\tau_c\,|\calN|\,(|\calN|-1)/2 \leq O\mleft(\tau_c\,|\calN|^3\mright).
\end{align}
Otherwise, the worst-case $\calG_{dir}$ should have $v_l$ being the first vertex of $p$ that is adjacent to a vertex $\bar{v}\in p$,
Then we have
{\allowdisplaybreaks\begin{align}
    &\max_{\calG_{dir}}T_{min}(\calG_{dir}) \nonumber\\
    &= \sum_{i\,=\,1}^{l-1} i\tau_c \times d(i,i+1) + \sum_{i\,=\,l}^{|\calN|-1} i\tau_c \times d(i,i+1) \nonumber\\\label{aux31:1}
    &\leq \sum_{i\,=\,1}^{l-1} i\tau_c\times 1 + \sum_{i\,=\,l}^{|\calN|-1} i\tau_c \times (l-1) \\\label{aux31:2}
    &=\frac{1}{2}\tau_c\,[l(l-1) + (l-1)(|\calN|+l-1)(|\calN|-l)] \\\nonumber
    &= O\mleft(\tau_c\,|\calN|^3\mright), 
\end{align}}where \cref{aux31:1} holds since no path in $\calG_{dir}$ is longer than $p$, and the maximum of \cref{aux31:2} is taken when $l = \lceil{\sqrt{3|\calN|^2-3|\calN|+3}/3}\rceil$. 
In all, \dfs has a $O\mleft(\tau_c\,|\calN|^3\mright)$ worst-case communication time. \qed

\section{Guarantees on Approximation Performance and Decision Time of \scenario{DFS-BSG}}
\begin{theorem}[Approximation Performance of \scenario{DFS-BSG}]\label{th:dfs-bsg-suboptimality}
\scenario{DFS-BSG} enjoys the suboptimality performance,
\begin{equation}\label{eq:dfs-bsg-suboptimality}
        \mathbb{E}\mleft[\frac{1}{T}\sum_{t=1}^T f(\calA_t^{{\fontsize{9}{9}\selectfont\sf DFS\text{-}BSG}})\mright] \geq \frac{1}{1+\curv_f}\,f(\solopt)- \Tilde{O}\mleft(|\calN|\sqrt{\frac{|\bar{\calV}|}{T}}\mright).
\end{equation}
\end{theorem}
\paragraph*{Proof} Eq.~\eqref{eq:dfs-bsg-suboptimality} holds by replacing the bound of \scenario{Exp3$^\star$-SIX} in~\cite[Theorem 2]{xu2023bandit} with that of \scenario{Exp3}~\cite{auer2002nonstochastic}. \qed

\begin{theorem}[Convergence Time of \scenario{DFS-BSG}]\label{th:dfs-bsg-decision-time} 
\scenario{DFS-BSG} requires $O\mleft[(\tau_f\,|\bar{\calV}|\,|\calN|^2\,+\,\tau_c\,|\calN|^5)\,|\bar{\calV}|\,/\,\epsilon\mright]$ time to converge in a directed network, and $O\mleft[(\tau_f\,|\bar{\calV}|\,|\calN|^2\,+\,\tau_c\,|\calN|^4)\,|\bar{\calV}|\,/\,\epsilon\mright]$ time in an undirected network.
\end{theorem} 

\paragraph*{Proof} Since the algorithm requires $O\mleft(|\bar{\calV}|\,|\calN|^2\,/\,\epsilon\mright)$ rounds to converge per \cref{eq:dfs-bsg-suboptimality}, and that in the worst case, each round requires $O\mleft(\tau_f\,|\bar{\calV}|\,+\,\tau_c\,|\calN|^3\mright)$ time for a directed network and $O\mleft(\tau_f\,|\bar{\calV}|\,+\,\tau_c\,|\calN|^2\mright)$ time for an undirected network, \Cref{th:dfs-bsg-decision-time} holds. \qed

\section{Comparison of \neighborsel with Nearest and Random Neighbor Selection: From Sparse to Dense Networks}
We also compare \alg with the heuristic benchmarks for neighbor selection (\textit{Nearest Neighbors} and \textit{Random Neighbors}) in 10 scenarios with different network densities. We adjust the network density by deploying the identical 20 cameras to cover areas of 10 different sizes. The results are presented in Fig.~\ref{fig:comparison-neighbor-selection}, where we observe a consistent improvement of \alg across all tested network densities. 

\myParagraph{Setup} %
    \textit{Environment:} The environment is static and square with areas ranging in $\{200, 400, \dots, 2000\}$.
    \textit{Agents:} There exist $20$ cameras. In each trial, the location $x_{i}$ of each camera $i\in\calN$ is uniformly sampled on the map. They all have the same communication range $c_i=16$. 
    \textit{Actions:} All cameras $i\in\calN$ have an FOV of radius $r_i=8$ and AOV $\theta_i=\pi/3$, with direction $a_{i,t}$ chosen from the 16 cardinal directions $\calV_i$, $\forall t$. 
    \textit{Objective Function:}  $f(\{a_{i,t}\}_{i\in \calN})$ is the total area of interest covered by the cameras $\calN$ when they select $\{a_{i,t}\}_{i\in \calN}$ as their FOV directions. $f$ is proved to be submodular~\cite{corah2018distributed}.

\myParagraph{Performance Metrics} We evaluate the achieved objective value of the algorithms over 2000 decision rounds.

\myParagraph{Results} We observe the improved coordination performance provided by network optimization via \neighborsel (Fig.~\ref{fig:comparison-neighbor-selection}). In particular, \alg is comparable or better to the two heuristic benchmarks across all presented network densities. The performance gaps compared to the benchmarks first increase then decrease as the network becomes denser. The reason is that when the network is very sparse, all potential neighbors $\calM_i$ are not informative enough to make a difference; when the network is too dense, multiple potential neighbors in $\calM_i$ can be informative enough and thus all strategies perform similarly well.

\bibliographystyle{IEEEtran}
\bibliography{references}

\begin{IEEEbiography}[{\includegraphics[width=1in,height=1.25in,keepaspectratio]{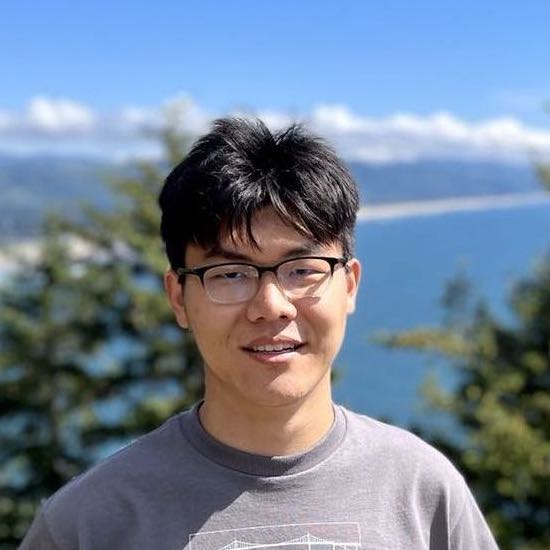}}]{Zirui Xu} (Graduate Student Member, IEEE) received the B.Eng. degree in Automation from Northeastern University, Shenyang, China, in 2018, and the M.S. degree in Electrical and Computer Engineering from Georgia Institute of Technology, Atlanta, GA, USA, in 2020. He is currently pursuing the Ph.D. degree with the Department of Aerospace Engineering, University of Michigan, Ann Arbor, MI, USA. His research focuses on theories, algorithms, and systems for scalable and reliable coordination of distributed multi-robot systems in resource-constrained, unstructured, and untrustworthy environments. 
\end{IEEEbiography}

\begin{IEEEbiography}[{\includegraphics[width=1in,height=1.25in,keepaspectratio]{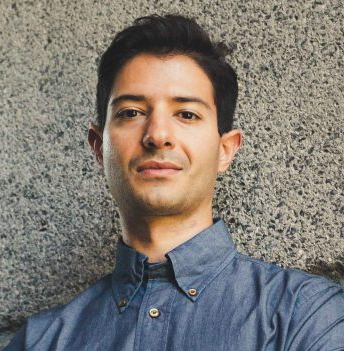}}]
{Vasileios Tzoumas} (Senior Member, IEEE) received his Ph.D. in Electrical and Systems Engineering at the University of Pennsylvania (2018). He holds a Master of Arts in Statistics from the Wharton School of Business at the University of Pennsylvania (2016), a Master of Science in Electrical Engineering from the University of Pennsylvania (2016), and a diploma in Electrical and Computer Engineering from the National Technical University of Athens (2012). Vasileios is an Assistant Professor in the Department of Aerospace Engineering, University of Michigan, Ann Arbor. Previously, he was at the Massachusetts Institute of Technology (MIT), in the Department of Aeronautics and Astronautics, and in the Laboratory for Information and Decision Systems (LIDS), where he was a research scientist (2019-2020) and a post-doctoral associate (2018-2019). Vasileios works on algorithms and innovative hardware for scalable and reliable cyber-physical systems in resource-constrained, uncertain, and contested environments via resource-aware decision-making, online learning, and resilient adaptation. 
Vasileios is a recipient of an NSF CAREER Award, an Army Research Office Early Career Program (ECP) Award, the Best Paper Award in Robot Vision at the 2020 IEEE International Conference on Robotics and Automation (ICRA), an Honorable Mention from the 2020 IEEE Robotics and Automation Letters (RA-L), and was a Best Student Paper Award finalist at the 2017 IEEE Conference on Decision and Control (CDC).
\end{IEEEbiography}

\end{document}